\documentclass[lettersize,journal]{IEEEtran}
\usepackage{amsmath,amsfonts}
\usepackage{algorithmic}
\usepackage{algorithm}
\usepackage{array}
\usepackage[caption=false,font=normalsize,labelfont=sf,textfont=sf]{subfig}
\usepackage{textcomp}
\usepackage{stfloats}
\usepackage{url}
\usepackage{verbatim}
\usepackage{graphicx}
\usepackage{cite}
\usepackage{multirow}
\usepackage{booktabs}
\begin{document}

\title{A Novel Learning-based Robust Model Predictive \\Control Energy Management Strategy \\for Fuel Cell Electric Vehicles}

\author{Shibo Li, Zhuoran Hou, Liang Chu, Jingjing Jiang, Yuanjian Zhang,~\IEEEmembership{Member,~IEEE}
\thanks{ \emph{(Corresponding author: Yuanjian Zhang.)}

Shibo Li, Zhuoran Hou, and Liang Chu are with the School of College of Automotive Engineering, Jilin University, Changchun 130022, China (e-mail: lisb22@mails.jlu.edu.cn; houzr20@mails.jlu.edu.cn; chuliang@jlu.edu.cn)

Jingjing Jiang and Yuanjian Zhang are with the Department of Aeronautical and Automotive Engineering, Loughborough University, Loughborough, U.K (e-mail: j.jiang2@lboro.ac.uk; y.y.zhang@lboro.ac.uk)
}}

\markboth{Journal of \LaTeX\ Class Files,~Vol.~14, No.~8, August~2021}%
{Shell \MakeLowercase{\textit{et al.}}: A Sample Article Using IEEEtran.cls for IEEE Journals}

\IEEEpubid{0000--0000/00\$00.00~\copyright~2021 IEEE}

\maketitle

\begin{abstract}
The multi-source electromechanical coupling makes the energy management of fuel cell electric vehicles (FCEVs) relatively nonlinear and complex especially in the types of 4-wheel-drive (4WD) FCEVs. Accurate state observing for complicated nonlinear system is the basis for fantastic energy managing in FCEVs. Aiming at releasing the energy-saving potential of FCEVs, a novel learning-based robust model predictive control (LRMPC) strategy is proposed for a 4WD FCEV, contributing to suitable power distribution among multiple energy sources. The well-designed strategy based on machine learning (ML) translates the knowledge of the nonlinear system to the explicit controlling scheme with superior robust performance. To start with, ML methods with high regression accuracy and superior generalization ability are trained offline to establish the precise state observer for SOC. Then, explicit data tables for SOC generated by state observer are used for grabbing accurate state changing, whose input features include the vehicle status and the states of vehicle components. To be specific, the vehicle velocity estimation for providing future speed reference is constructed by deep forest. Next, the components including explicit data tables and vehicle velocity estimation are combined with model predictive control (MPC) to release the state-of-the-art energy-saving ability for the multi-freedom system in FCEVs, whose name is LRMPC. At last, the detailed assessment is performed in simulation test to validate the advancing performance of LRMPC. The corresponding results highlight the optimal control effect in energy-saving potential and strong real-time application ability of LRMPC.
\end{abstract}


\section{Introduction}
\IEEEPARstart{N}{owadays,} fuel cell electric vehicles (FCEVs) are increasingly favored by the automotive industry due to using of clean energy and mitigating global warming \cite{ref1, ref2}. Tailored powertrain configurations have given FCEVs extraordinary performance to effortlessly balance driving range, energy storage, and drivability \cite{ref3, ref4, ref5}. As a complex system with multiple energy and power degrees of freedom, 4WD FCEVs urgently require a rationally designed energy management strategy (EMS) to optimize the vehicle energy flow distribution to release the energy-saving potential \cite{ref6, ref7}. However, dealing with such highly nonlinear and complex systems remains a pressing challenge.

EMSs for FCEVs can be divided into two categories: rule-based EMSs and optimization-based EMSs \cite{ref8, ref9}. Rule-based strategies are usually derived from the engineering knowledge of experts without considering the prior knowledge of driving cycles, such as state machine control \cite{ref10} and fuzzy logic control \cite{ref11}. The excellent real-time ability and reliability \cite{ref12, ref13} make rule-based strategies extremely easy to implement, but the accurate formulation of rules is difficult \cite{ref14, ref15} which can directly influence the control trajectory. As a result, it is not guaranteed to provide adaptive control based on changes in the future driving conditions. Optimization-based strategies can be subdivided into global optimization strategies and instantaneous optimization strategies respectively \cite{ref16, ref17}. Global optimization methods such as dynamic programming (DP) \cite{ref18} and pontryagin’s minimum principle (PMP) \cite{ref19} are required to obtain the global optimal control trajectory by prior knowledge of the whole driving cycles in advance. Besides, global optimization methods are plagued by the "curse of dimensionality" with large computational consumption of calculating resources \cite{ref20, ref21, ref22}, leading to being implemented offline as a baseline reference. In contrast, instantaneous optimization strategies require only partial pre-knowledge of short-time future driving conditions to solve the local optimal control trajectory at a certain instant, which have received much attention from researchers. Furthermore, as one of the instantaneous optimization strategies, the equivalent consumption minimization strategy (ECMS) can solve the current optimal multi-power source power distribution at a certain instantaneous moment to obtain the optimal performance of energy consumption \cite{ref23, ref24}. Compared with ECMS, model predictive control (MPC) is more energy-efficient because of obtaining the global optimal solution in the predicted domain \cite{ref25}. Therefore, the local optimal control characteristics of MPC can give FCEVs a better energy-saving potential.
\IEEEpubidadjcol
Nevertheless, MPC still has a few problems to be solved in practical application. The state observation model of MPC, reflecting the state of power system components \cite{ref26}, plays an important role in the control process. On this basis, improving the accuracy of the state observation model becomes an essential approach to optimize MPC \cite{ref27, ref28}. During the process of application, traditional MPC (T-MPC) expresses the state changes of vehicle power system components in the form of simplified mathematical models, such as the first-order resistance-capacitance (RC) model \cite{ref29, ref30} and the coulomb counting method \cite{ref31, ref32}, to perform real-time observation of the state vector. Although the above models are simple and fast to calculate, the charge and discharge process of batteries is highly nonlinear and specific due to the influence of the environment. Heavy reliance on a model makes T-MPC susceptible to modeling error and external disturbances, leading to poor performance or instability. 

To enhance the control effect, robust MPCs \cite{ref33, ref34} can improve the robustness of the state observation model by external correction, aiming to more accurately represent the state of power system components. Moreover, the robust MPC based on the min-max game \cite{ref35} and the robust tube-based MPC \cite{ref36} have shown extraordinary control superiority in this field. The robust MPC based on the min-max game belongs to a modified control method considering the deviations caused by system uncertainty. However, it is worth noting that the min-max game framework is relatively conservative and the control trajectory is even not solvable. This phenomenon is caused by the goal that deals with the worst case in the complete set of the state of power system components. As for the robust tube-based MPC, it decomposes the robust MPC into an offline robust controller design and an online open-loop MPC problem and improves the control performance by optimizing over control policies. Unfortunately, the error convergence ability of the partition in which the current state lies will directly influence the performance. Therefore, the above two types of robust MPCs can only enhance the control performance by externally correcting the control trajectory rather than the internal state observation model, leading to difficult solving and dissatisfactory error convergence ability.

As another solution, in the field of artificial intelligence (AI), machine learning (ML) methods such as supervised learning \cite{ref37}, unsupervised learning \cite{ref15} and semi-supervised learning \cite{ref38, ref39} have played an important role in the design of energy management strategies due to strong regression and classification capabilities, especially in powertrain modeling \cite{ref40}, reference trajectory generation \cite{ref22}, pattern recognition \cite{ref41}, etc. Specifically, the randomness of the bootstrap sample and threshold formation makes the model less susceptible to overfitting and robust to external noise, which give machine learning models superior generalization ability, high accuracy and interpretability \cite{ref42}. These advantages mentioned above are extremely suitable for the construction of the state observer, which reflects the state observation model. First, the superior generalization ability can improve the observation instability caused by the changing of state observer with the driving conditions. Second, the high accuracy can reduce the state observation error due to the simplified mathematical model. Third, the strong interpretability can help the state observer to interpret and deduce the state changes. Therefore, the combination of machine learning methods and state observer based on MPC is a reasonable solution to the existing problem of robust MPC. However, even though machine learning can enhance online application ability of MPC through offline training and online deployment, the relatively complex mapping relationship will restrict the ability of real-time application. As a result, it is necessary to establish a state observer with high accuracy and strong interpretability and improve the feasibility of online application based on machine learning methods in the on-board controller.

On this account, this paper presents a novel learning-based robust model predictive control (LRMPC) energy management strategy based on ML methods for a 4WD FCEV. In the designed LRMPC, the state observer integrated into LRMPC can reflect accurate changing of SOC based on the basic ML methods. The basic ML models are elaborately trained by extracting various component characteristics aiming at guarantee the controlling effect. To furnish LRMPC with distinctive capabilities in energy management knowledge interpretation, the explicit data tables are obtained based on ML models strengthening the efficiency of the novel strategy. And then, deep forest (DF) is adopted to obtain the precise future driving velocity in real-time driving. The simulation results validate the advancement of the proposed strategy.

The contributions are provided in: 
\begin{enumerate}
\item{A learning-based robust model predictive control energy management control framework is proposed for 4WD FCEVs with a better control performance by representing the nonlinear characteristics of the vehicle power system components through multiple variables.}
\item{A data-driven velocity estimation method based on deep forest is proposed with a competitive velocity prediction error considering complex driving conditions.}
\item{The superior accuracy and robustness of LRMPC in state observation are verified by comparing the performance of five machine learning methods and traditional MPC over different prediction domains, including energy-saving effects and real-time performances.}
\item{Explicit solutions for machine learning models are integrated into the MPC controller, resulting in improving real-time application capabilities.}
\end{enumerate}

The rest of this paper is organized as follows: section 2 presents the modeling of the FCEV. In Section 3, the vehicle velocity estimation method based on deep forest and the LRMPC energy management strategy are proposed. Simulation results and comparative analysis are described in Section 4. Finally, conclusions are drawn in Section 5.

\section{Powertrain configuration \\and system modeling}
\subsection{Powertrain configuration}
\begin{figure}[!t]
\centering
\includegraphics[width=3.2in]{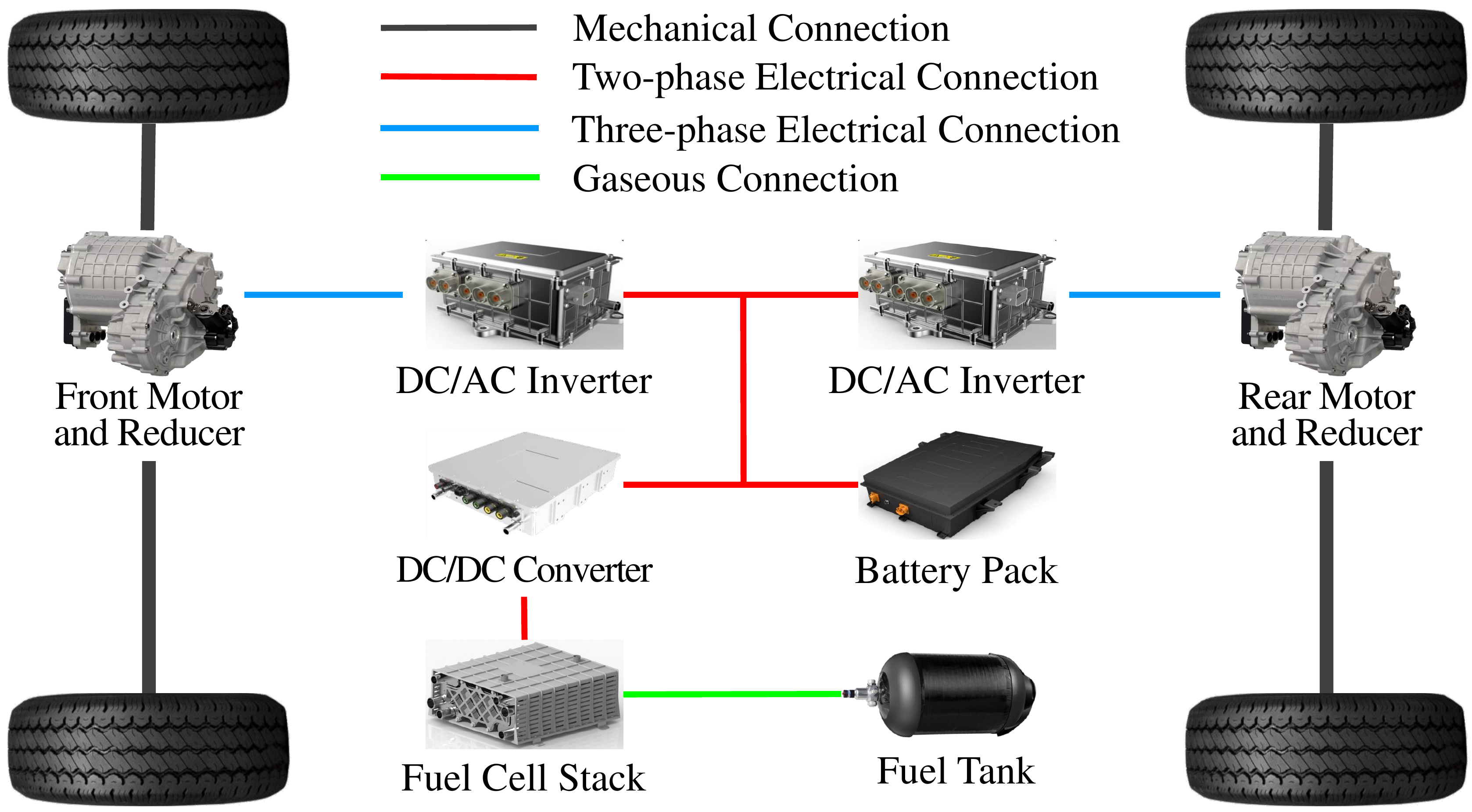}
\caption{The powertrain configuration of the FCEV.}
\label{F:Fig1}
\end{figure}

As shown in Fig. \ref{F:Fig1}, in the powertrain of the FCEV studied in this paper, a unidirectional DC/DC connects the fuel cell system to the DC bus, forming a dual energy source together with the battery pack. The front and rear motors are respectively supplied with energy via DC/AC inverters to drive the vehicle. Note that the power of the front and rear motors is distributed according to a fixed distribution ratio, which is numbered of 0.6 and 0.4 respectively. Besides, the driving mode of the studied FCEV is divided into pure electric vehicle (EV) mode and hybrid electric vehicle (HEV) mode. In HEV mode, EMS provides power distribution for fuel cell system and battery. The basic parameters of the FCEV are shown in TABLE \ref{T:Table1}.

\begin{table}[!t]
\caption{Basic parameters of the FCEV\label{T:Table1}}
\centering
\setlength{\tabcolsep}{1mm}{
\begin{tabular}{c c}
\toprule
Characteristic & Value\\
\hline
Mass & 1860kg\\
Tire rolling radius & 0.35m\\
Rolling resistance coefficient & 0.015\\
Aerodynamic drag coefficient & 0.3\\
Frontal area & 2m$^{2}$\\
Air density & 1.18kg/m$^{3}$\\
Front/Rear motor speed range & 0-14000rpm/0-10000rpm\\
Front/Rear motor torque range & -137-137Nm/-195-195Nm\\
Battery capacity & 40Ah\\
Maximum power of FC stack & 61.56kW\\
DC/DC average efficiency & 87.75$\%$\\
\bottomrule 
\end{tabular}}
\end{table}

Assuming that the vehicle is driven on a flat road with sufficient tire-road friction coefficient, the vehicle longitudinal dynamics equation can be defined as:
\begin{equation}\label{E:Equ1}
{{F}_{t}}(t)=mgf+\frac{1}{2}{{C}_{D}}AV_{t}^{2}(t)+\delta m\frac{d{{V}_{t}}}{dt}
\end{equation}
where ${{F}_{t}}$ is the vehicle driving force, $m$ is the vehicle mass, $g$ is the acceleration of gravity, $f$ is the rolling resistance coefficient, ${{C}_{D}}$ is the aerodynamic drag coefficient, $A$ is the vehicle frontal area, ${{V}_{t}}(t)$ is the vehicle velocity, and $\delta$ is the weighting factor of the rotating mass. Similarly, the equilibrium equation of vehicle driving force and driving obstruction force is as follows:
\begin{equation}\label{E:Equ2}
{{P}_{t}}(t)=\left( mgf+\frac{1}{2}{{C}_{D}}AV_{t}^{2}(t)+\delta m\frac{d{{V}_{t}}}{dt} \right){{V}_{t}}(t)
\end{equation}

Therefore, the vehicle demand power used to drive can be defined as:
\begin{equation}\label{E:Equ3}
{{P}_{load}}(t)=\frac{{{P}_{t}}(t)}{{{\eta }_{DC/AC}}{{\eta }_{motor}}}
\end{equation}
where ${{\eta}_{DC/AC}}$ and ${{\eta}_{motor}}$ are the DC/AC efficiency and motor efficiency, respectively. Obviously, the power of the dual energy source should satisfy the following equation:
\begin{equation}\label{E:Equ4}
{{P}_{load}}(t)={{P}_{fc}}(t)+{{P}_{batt}}(t)
\end{equation}
where ${{P}_{fc}}$ is the output net power of the fuel cell system and ${{P}_{batt}}$ is the battery power.

\subsection{Fuel cell stack model}
\begin{figure}[!t]
\centering
\includegraphics[width=3in]{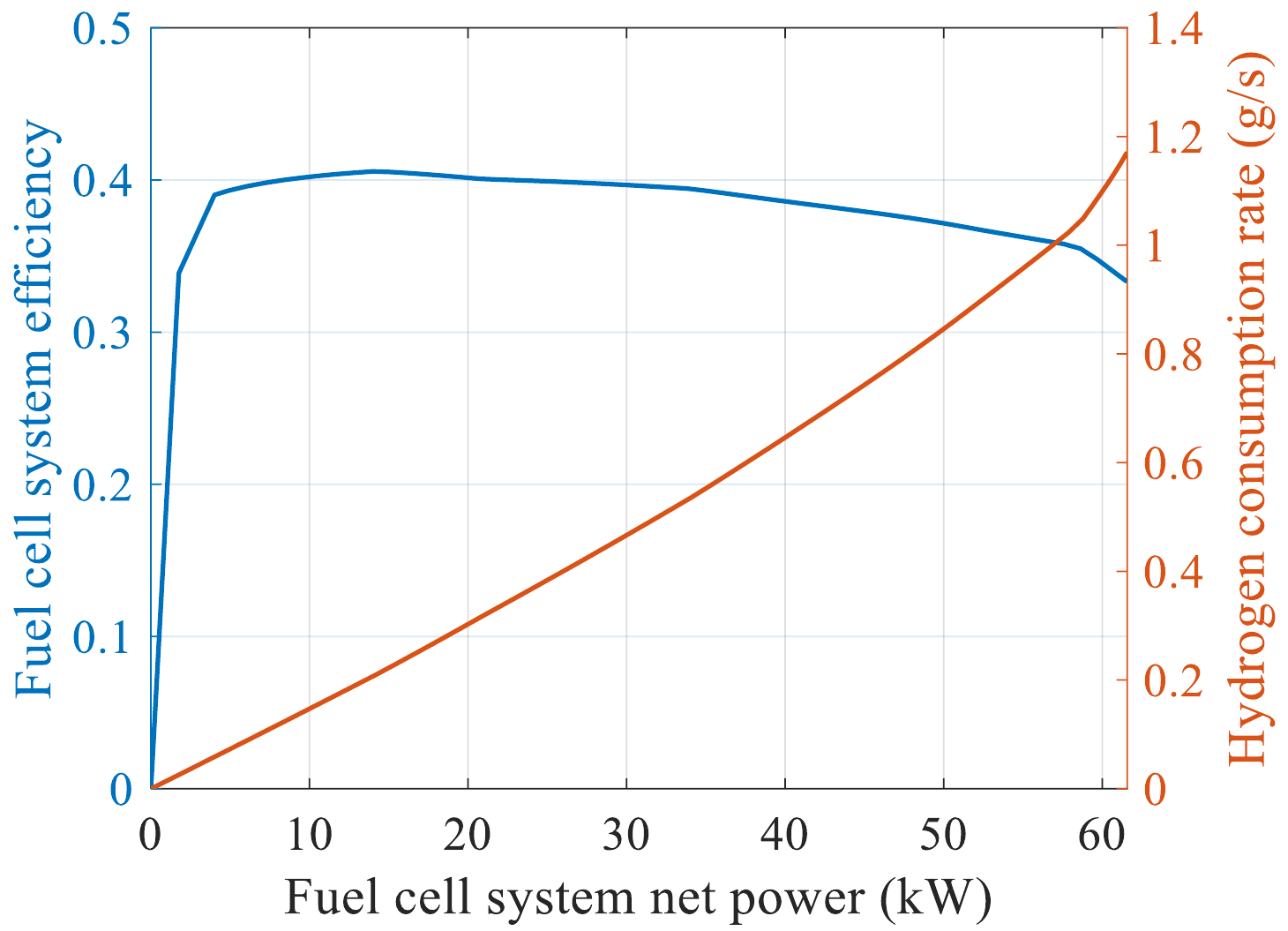}
\caption{Fuel cell system model.}
\label{F:Fig2}
\end{figure}

The FCEV studied in this paper applies a proton exchange membrane fuel cell (PEMFC). The static model of this fuel cell system is obtained based on experimental data assuming that the fuel cell ambient temperature, moderation, and gas pressure are maintained constant. And the difference in gas pressure between the anode and cathode is neglected. The relationship between power and efficiency and the relationship between power and hydrogen consumption rate of the fuel cell system are shown in Fig. \ref{F:Fig2}.

Theoretically, the hydrogen consumption of the fuel cell system is related to the output power as:
\begin{equation}\label{E:Equ5}
{{m}_{{{H}_{2}}}}=\frac{1}{LH{{V}_{{{H}_{2}}}}}\int{\frac{{{P}_{fc}}}{{{\eta }_{fc}}({{P}_{fc}})}}
\end{equation}
where ${{m}_{{{H}_{2}}}}$ is the hydrogen mass, $LH{{V}_{{{H}_{2}}}}$ is the hydrogen low calorific value, here taken as 120MJ/kg, ${{P}_{fc}}$ is the output power, and ${{\eta}_{fc}}$ is the stack efficiency. In this study, the calculation of hydrogen consumption is based on the fuel cell power obtained by looking up the above table as shown in Fig. \ref{F:Fig2} and integrating:
\begin{equation}\label{E:Equ6}
{{m}_{{{H}_{2}}}}=\int{{{{\dot{m}}}_{{{H}_{2}}}}({{P}_{fc}})}
\end{equation}

\subsection{Battery model}
Compared with supercapacitors, batteries have obvious advantages of high energy density, but the charging and discharging performance of batteries are directly influenced by the internal resistance. In this study, the battery is modeled using an equivalent circuit model considering the internal resistance. Furthermore, both the open circuit voltage and internal resistance of the battery are nonlinear related to the charge state and temperature, which is obtained from experimental data. Therefore, the battery pack terminal voltage can be written as a function of the open circuit voltage and internal resistance:
\begin{equation}\label{E:Equ7}
{{U}_{batt}}={{U}_{OCV}}-{{R}_{batt}}{{I}_{batt}}
\end{equation}
where ${{U}_{batt}}$ is the pack terminal voltage, ${{U}_{OCV}}$ is the pack open circuit voltage, ${{R}_{batt}}$ is the pack internal resistance, and ${{I}_{batt}}$ is the pack current. The relationship between pack current and power is as follows:
\begin{equation}\label{E:Equ8}
{{I}_{batt}}=\frac{{{U}_{OCV}}-\sqrt{U_{OCV}^{2}-4{{R}_{batt}}{{P}_{batt}}}}{2{{R}_{batt}}}
\end{equation}
where ${{P}_{batt}}$ is the pack power, and define ${{P}_{batt}}>0$ when the battery is discharged. According to the coulomb counting method, the battery SOC can be described as:
\begin{equation}\label{E:Equ9}
SOC(t)=SOC({{t}_{0}})-\frac{\int_{{{t}_{0}}}^{t}{{{I}_{batt}}dt}}{3600{{C}_{batt}}}
\end{equation}

\subsection{Front and rear motor model}
\begin{figure*}[!t]
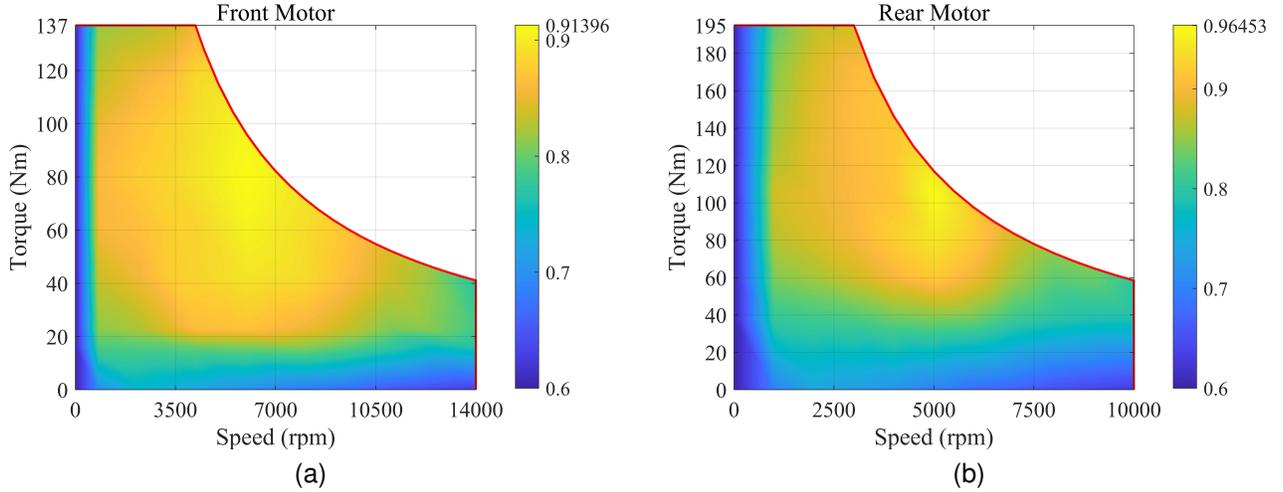

\centering
\subfloat[]{\includegraphics[width=3.2in]{Fig3a}%
\label{F:Fig3_1}}
\hfil
\subfloat[]{\includegraphics[width=3.2in]{Fig3b}%
\label{F:Fig3_2}}
\caption{Front and rear motor model.}
\label{F:Fig3}
\end{figure*}

In the studied FCEV, permanent magnet synchronous AC motors with peak torque of 137Nm and 195Nm are applied to the front and rear axles respectively, and their efficiency is a function of motor torque and speed:
\begin{equation}\label{E:Equ10}
{{\eta }_{motor}}=f({{T}_{motor}},{{n}_{motor}})
\end{equation}
where ${{T}_{motor}}$ is the motor torque, and ${{n}_{motor}}$ is the motor speed. The efficiency of the motor is obtained by looking up characteristic maps shown in Fig. \ref{F:Fig3}.

\section{Energy management strategy based on LRMPC}
\begin{figure*}[!t]
\centering
\includegraphics[width=5.5in]{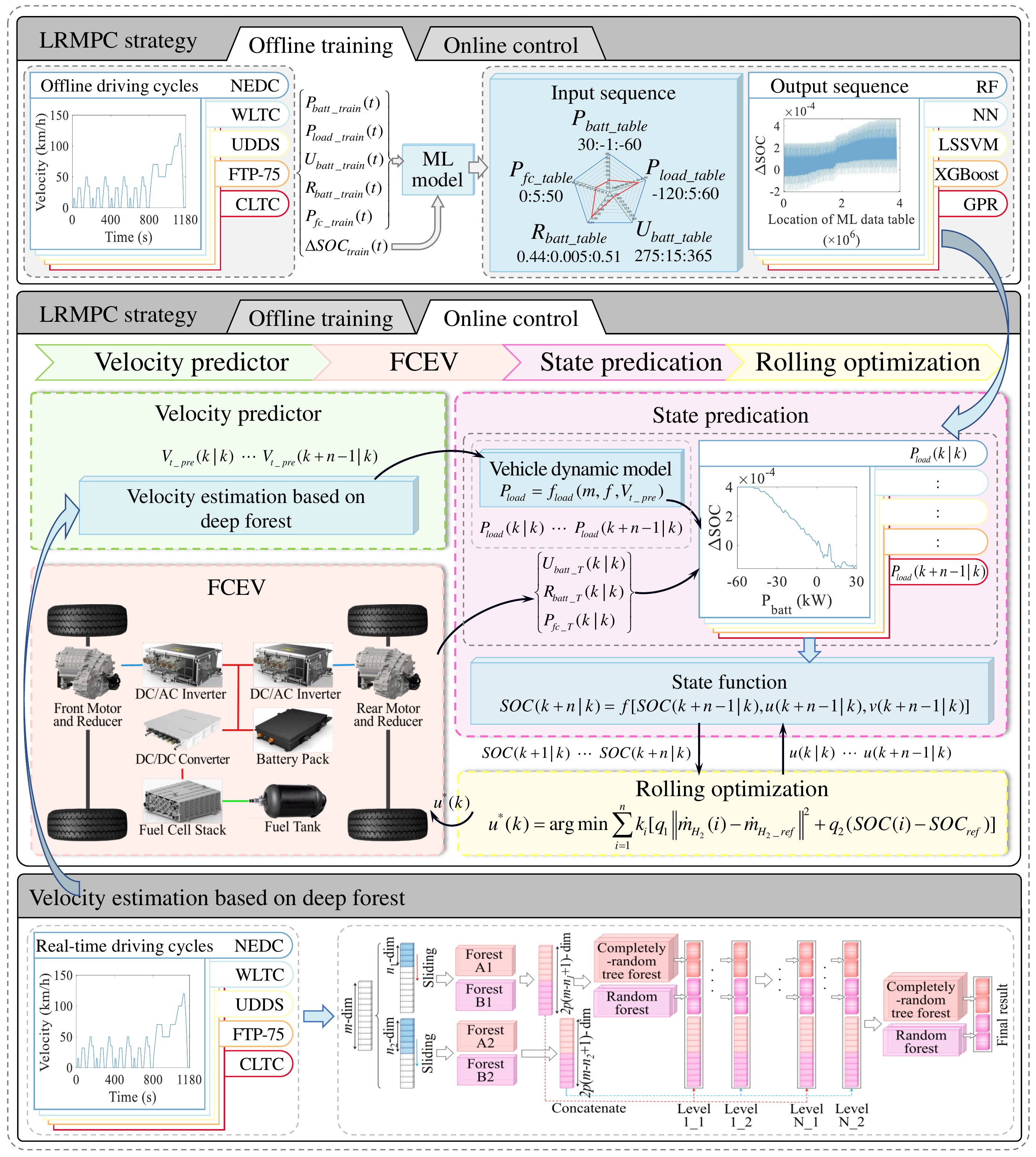}
\caption{Energy management strategy based on LRMPC.}
\label{F:Fig4}
\end{figure*}

In this section, the LRMPC energy management control framework is divided into two parts including offline training and online control, as shown in Fig. \ref{F:Fig4}. In the part of offline training, the state observation models are trained based on five basic ML methods and then integrated into the module of state prediction in the part of online control, aiming at accurately enhancing the performance of state observation. As for the part of online control, the vehicle velocity estimation method based on the deep forest is proposed to generate the reference trajectory which is reflected in the power demand of the vehicle for MPC control. And then, both reference and feedback information of the FCEV is jointly used as inputs for state observation to obtain the optimal control sequence during MPC rolling optimization.

\subsection{Traditional MPC}
The control optimization problem based on MPC should contain a prediction model describing the control variables and state variables of the control system, which can obtain the optimal control trajectory and future information according to the current information and the future control variables in the field of vehicle control. Therefore, the prediction model can be expressed by the following state equation and observation equation:
\begin{equation}\label{E:Equ11}
\dot{x}=f(x,u,v)
\end{equation}
\begin{equation}\label{E:Equ12}
y=g(x,u,v)
\end{equation}
where $x$, $y$, $u$, and $v$ are the state variables, observation variables, control variables, and disturbance variables respectively.\\
Traditional MPC further derive specific expressions for the above equations through linearization and discretization, and the linearized equations of the prediction model are as follows:
\begin{equation}\label{E:Equ13}
\dot{x}=Ax+Bu+Cv
\end{equation}
\begin{equation}\label{E:Equ14}
y=Dx+Eu+Fv+G
\end{equation}
where $A$, $B$, and $C$ are the state transfer matrices and $D$, $E$, $F$, and $G$ are the observation matrices, obtained from the following relations:
\begin{equation}\label{E:Equ15}
\left \{\begin{aligned}
 A(k)&=\frac{\partial f}{\partial x}{{|}_{x=x(k),u=u(k),v=v(k)}}\\ 
 B(k)&=\frac{\partial f}{\partial u}{{|}_{x=x(k),u=u(k),v=v(k)}}\\ 
 C(k)&=\frac{\partial f}{\partial v}{{|}_{x=x(k),u=u(k),v=v(k)}}\\ 
 D(k)&=\frac{\partial g}{\partial x}{{|}_{x=x(k),u=u(k),v=v(k)}}\\ 
 E(k)&=\frac{\partial g}{\partial u}{{|}_{x=x(k),u=u(k),v=v(k)}}\\ 
 F(k)&=\frac{\partial g}{\partial v}{{|}_{x=x(k),u=u(k),v=v(k)}}\\ 
 G(k)&=g(x(k),u(k),v(k))-D(k)x(k) \\ 
 &+E(k)u(k)+F(k)v(k)
\end{aligned}
\right.
\end{equation}

To obtain the single-step state transfer results in the prediction domain, the discretization equation of the state observation model can be expressed as:
\begin{equation}\label{E:Equ16}
\left\{ \begin{aligned}
  & x(k+1|k)=x(k|k)+\dot{x}(k|k)\cdot T \\ 
 & \text{      }: \\ 
 & x(k+n|k)=x(k+n-1|k)+\dot{x}(k+n-1|k)\cdot T  
\end{aligned} \right.
\end{equation}

In traditional MPC, the amount of state transfer can be expressed as:
\begin{equation}\label{E:Equ17}
\left\{ \begin{aligned}
  & \dot{x}(k|k)=A(k)x(k|k)+B(k)u(k|k)+C(k)v(k|k) \\ 
 & \text{      }: \\ 
 & \dot{x}(k+n-1|k)=A(k+n-1)x(k+n-1|k) \\ 
 & \ \ \ \ \ \ \ \ \ \ \ \ \ \ \ \ \ \ \ \ \ +B(k+n-1)u(k+n-1|k) \\ 
 & \ \ \ \ \ \ \ \ \ \ \ \ \ \ \ \ \ \ \ \ \ +C(k+n-1)v(k+n-1|k) \\ 
\end{aligned} \right.
\end{equation}

Therefore, the final form of the state observation model discretization equation is as follows:
\begin{equation}\label{E:Equ18}
\tilde{X}=\tilde{A}x(k|k)+\tilde{B}\tilde{U}+\tilde{C}\tilde{V}
\end{equation}
\begin{equation}\label{E:Equ19}
\left \{\begin{aligned}
\tilde{X}&={{[x(k+1|k),\cdots ,x(k+n|k)]}^{T}} \\ 
\tilde{U}&={{[u(k|k),\cdots ,u(k+n-1|k)]}^{T}} \\ 
\tilde{V}&={{[v(k|k),\cdots ,v(k+n-1|k)]}^{T}} \\ 
\tilde{A}&={{[1,\cdots ,1]}^{T}} \\ 
\tilde{B}&=\left[ \begin{matrix}
   B(k) & 0 & \cdots  & 0  \\
   \vdots  & \ddots  & \ddots  & \vdots   \\
   B(k) & \cdots  & B(k+n-1) & 0  \\
\end{matrix} \right]\cdot T \\ 
\tilde{C}&=\left[ \begin{matrix}
   C(k) & 0 & \cdots  & 0  \\
   \vdots  & \ddots  & \ddots  & \vdots   \\
   C(k) & \cdots  & C(k+n-1) & 0  \\
\end{matrix} \right]\cdot T \\ 
\end{aligned}
\right.
\end{equation}

Based on the specific needs of the FCEV energy management problem, the prediction model can be built as follows:
\begin{equation}\label{E:Equ20}
S\dot{O}C=-\frac{{{U}_{OCV}}-\sqrt{U_{OCV}^{2}-4{{R}_{batt}}{{P}_{batt}}}}{2{{R}_{batt}}{{C}_{batt}}}
\end{equation}
\begin{equation}\label{E:Equ21}
{{\dot{m}}_{{{H}_{2}}}}={{\dot{m}}_{{{H}_{2}}\_fc}}+{{\dot{m}}_{{{H}_{2}}\_batt}}
\end{equation}
\begin{equation}\label{E:Equ22}
\left \{\begin{aligned}
{{{\dot{m}}}_{{{H}_{2}}\_fc}}&={{P}_{fc}}\cdot {{C}_{{{H}_{2}}}} \\ 
{{{\dot{m}}}_{{{H}_{2}}\_batt}}&=\frac{{{P}_{batt}}\cdot S}{LH{{V}_{{{H}_{2}}}}} \\ 
\end{aligned}
\right.
\end{equation}
where $SOC$ is the battery state of charge, ${{\dot{m}}_{{{H}_{2}}}}$ is the equivalent hydrogen consumption, ${{\dot{m}}_{{{H}_{2}}\_fc}}$ is the fuel cell hydrogen consumption, ${{\dot{m}}_{{{H}_{2}}\_batt}}$ is the battery hydrogen consumption, ${{C}_{{{H}_{2}}}}$ is the fuel cell hydrogen consumption rate, and $S$ is the equivalence factor.

In traditional MPC, define the state variable $x$, observation variable $y$, control variable $u$ and disturbance variable $v$ as follows:
\begin{equation}\label{E:Equ23}
x=SOC,y={{\dot{m}}_{{{H}_{2}}}},u=\frac{{{P}_{batt}}}{{{P}_{load}}},v={{P}_{load}}
\end{equation}

Therefore, the state transfer matrices and observation matrices satisfy the following equation:
\begin{equation}\label{E:Equ24}
\left \{\begin{aligned}
A(k)&=0,D(k)=0 \\ 
B(k)&=-\frac{{{P}_{load}}(k)}{{{C}_{batt}}}{{(U_{OCV}^{2}-4{{R}_{batt}}{{P}_{batt}}(k))}^{-0.5}} \\ 
C(k)&=-\frac{{{P}_{batt}}(k)}{{{P}_{load}}(k){{C}_{batt}}}{{(U_{OCV}^{2}-4{{R}_{batt}}{{P}_{batt}}(k))}^{-0.5}} \\ 
E(k)&=\left( -{{C}_{{{H}_{2}}}}+\frac{S}{Q} \right){{P}_{load}}(k) \\ 
F(k)&={{C}_{{{H}_{2}}}}+\left( -{{C}_{{{H}_{2}}}}+\frac{S}{Q} \right)\frac{{{P}_{batt}}(k)}{{{P}_{load}}(k)} \\ 
G(k)&=\left( {{C}_{{{H}_{2}}}}-\frac{S}{Q} \right){{P}_{batt}}(k)  
\end{aligned}
\right.
\end{equation}

To acquire the solution of optimal control by MPC, the nonlinear control problem can be described as:
\begin{equation}\label{E:Equ25}
\begin{aligned}
  & {{u}^{*}}(k)=\arg \min \sum\limits_{i=1}^{n}{{{k}_{i}}[{{q}_{1}}{{\left\| y(k+i|k)-{{y}_{ref}} \right\|}^{2}}} \\ 
 & \ \ \ \ \ \ \ \ \ \ \ \ \ \ \ \ \ \ \ \ \ \ \ \ \ \ \ \ +{{q}_{2}}(x(k+i|k)-{{x}_{ref}})] \\ 
\end{aligned}
\end{equation}

\begin{equation}
\label{E:Equ26}
\begin{aligned}
  & subject\text{ }to \\ 
 & \left\{ \begin{aligned}
  & u{{(k+i-1|k)}_{\min }}\le u(k+i-1|k)\le u{{(k+i-1|k)}_{\max }} \\ 
 & Mu(k+i-1|k)+N\ge 0 \\ 
 & {{P}_{batt\_\min }}\le {{P}_{batt}}(k+i-1|k)\le {{P}_{batt\_\max }} \\ 
 & {{P}_{fc\_\min }}\le {{P}_{fc}}(k+i-1|k)\le {{P}_{fc\_\max }} \\ 
 & \Delta {{P}_{fc\_\min }}\le \Delta {{P}_{fc}}(k+i-1|k)\le \Delta {{P}_{fc\_\max }} \\ 
\end{aligned} \right. \\ 
\end{aligned}
\end{equation}

In this study, the open source toolbox CasADi is used to solve the optimal control trajectory, where $u$, $x$, and $y$ are the control variables, state variables, and output variables, respectively. ${{y}_{ref}}$ and ${{x}_{ref}}$ are the tracking references. $M$ and $N$ are the constraint matrices. ${{k}_{i}}$ is a weight matrix, and ${{q}_{1}}$ and ${{q}_{2}}$ are the weight coefficients.

Since the conventional MPC uses linearized equations instead of the original state equations to achieve state observation. This linearization method adversely affects the accuracy of the state observation model, especially when the prediction domain increases, the cumulative effect of the error will enlarge the deviation of the state observation, resulting in a sub-optimal control result of MPC.

\subsection{LRMPC}
In order to meet the requirements of MPC state observation accuracy and real-time performance, the LRMPC strategy is proposed, which is divided into two major parts: offline training and online control, as shown in Fig. \ref{F:Fig4}. Compared with the traditional MPC, LRMPC can obtain offline training models using training datasets collected from the historical data of the FCEV based on ML methods. And then, explicit offline data tables for state observation are obtained by traversing the possible combinations of feature attribute inputs of the stare observation model using ML models. Further, the linearization method of traditional MPC state observation is replaced by the mapping $h$ characterized by the above data tables, improving the observation accuracy and real-time performance. The mapping $h$ is as follows:
\begin{equation}\label{E:Equ27}
\dot{x}=h(u,v,z)
\end{equation}
where $h$ is the mapping relationship of the state observation model, $z$ is the parameter vectors related to the mapping relationship, and $x$, $u$, and $v$ are the state variables, control variables, and disturbance variables respectively. Therefore, the discretization equation of the state observation model in LRMPC is as follows:
\begin{equation}\label{E:Equ28}
\left\{ \begin{aligned}
  & x(k+1|k)=x(k|k)+h(u(k|k),v(k|k),z(k|k))\cdot T \\ 
 & \text{      }: \\ 
 & x(k+n|k)=x(k+n-1|k)+h(u(k+n-1|k), \\ 
 & \ \ \ \ \ \ \ \ \ \ \ \ \ \ \ \ \ \ \ \ \ \ \ \ \ \ \ \ \ \ \ \ \ \ \ \ \ \ \ \ \ v(k+n-1|k), \\ 
 & \ \ \ \ \ \ \ \ \ \ \ \ \ \ \ \ \ \ \ \ \ \ \ \ \ \ \ \ \ \ \ \ \ \ \ \ \ \ \ \ \ z(k+n-1|k))\cdot T \\ 
\end{aligned} \right.
\end{equation}

Corresponding to the traditional MPC, the following variables are defined in LRMPC considering the energy management problem of this study:
\begin{equation}\label{E:Equ29}
\left \{\begin{aligned}
x&=SOC,y={{{\dot{m}}}_{{{H}_{2}}}},u={{P}_{batt}},v={{P}_{load}} \\ 
z&={{[{{U}_{batt}},{{R}_{batt}},{{P}_{fc}}]}^{T}} \\ 
\end{aligned}
\right.
\end{equation}

\begin{figure*}[!t]
\centering
\includegraphics[width=4.5in]{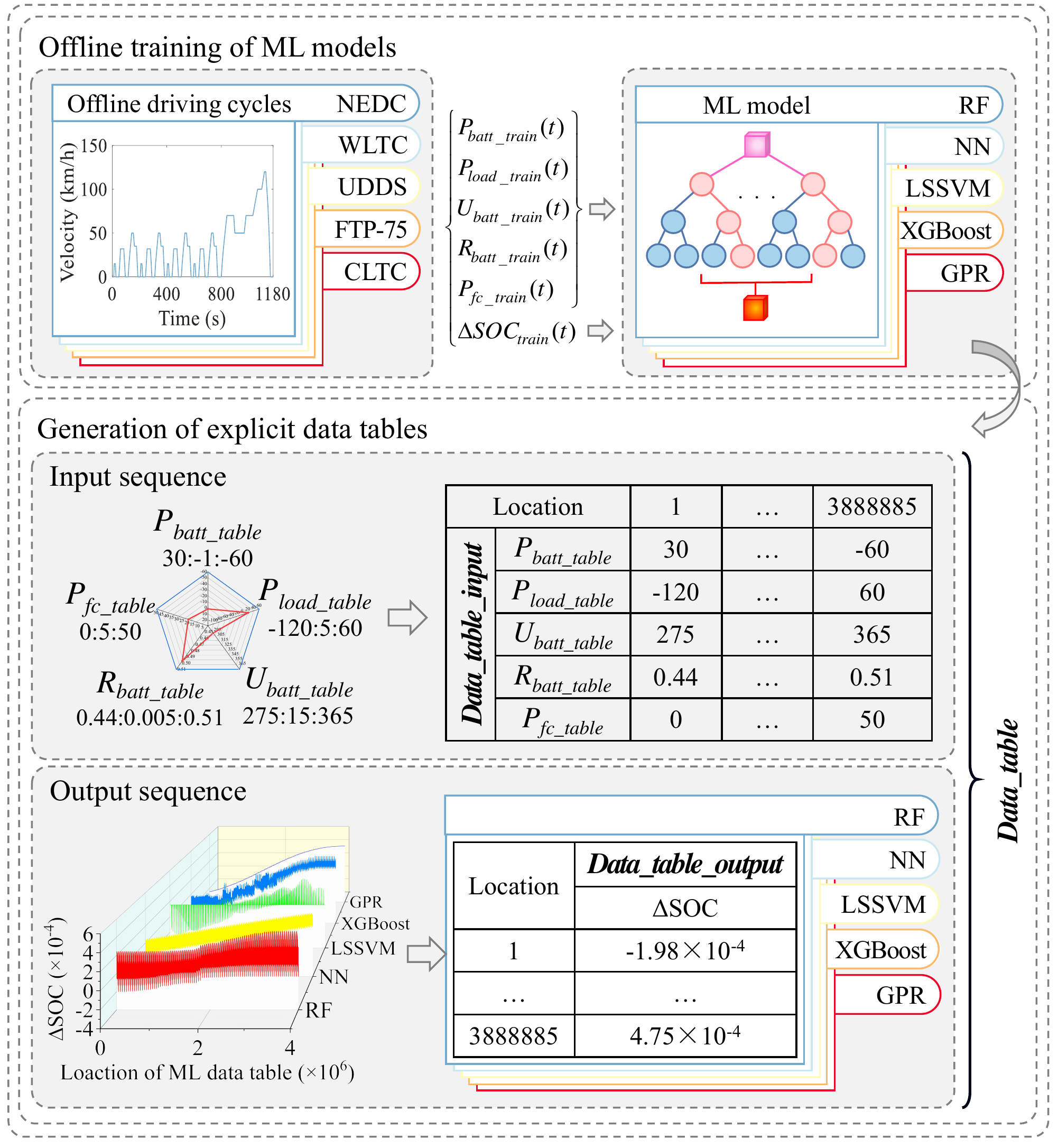}
\caption{The generation process of explicit data tables by offline training.}
\label{F:Fig5}
\end{figure*}

The explicit process of offline training of LRMPC provides guidance for obtaining the mapping relationship $h$ in the state observation model, as shown in Fig. \ref{F:Fig5}. In the first step of ML models offline training, the input feature sequences and output sequences are obtained through the historical data of the FCEV. The input sequence includes five feature attributes: battery power ${{P}_{batt\_train}}$, vehicle demand power ${{P}_{load\_train}}$, battery voltage ${{U}_{batt\_train}}$, battery internal resistance ${{R}_{batt\_train}}$ and fuel cell power ${{P}_{fc\_train}}$. The output sequence is the battery SOC variation $\Delta SO{{C}_{train}}$. Furthermore, the corresponding training models are obtained by five common ML methods: random forest (RF), neural network (NN), least square support vector machine (LSSVM), extreme gradient boosting (XGBoost), and gaussian process regression (GPR). In the second step of explicit data tables generating, considering the data range and scale of the state observation model, the five input features are discretized into a five-dimensional and 3888885 sets of input matrix $Data\_table\_input$ according to different ranges and intervals. And then, the corresponding output matrixes $Data\_table\_output$ are obtained based on the five ML models respectively. Thus, five explicit state observation data tables for online control are obtained, which are named $Data\_table$.

Algorithm \ref{A:Alg1} demonstrates the online control process of LRMPC. In contrast to the traditional MPC, LRMPC obtains the control and state variables from $Data\_table$ for the observation of state updates to replace the linearized equations in the traditional MPC. In this control process, considering the small total prediction domain under the simulation single step time of 0.05s, linearization is performed at time $k$, that is, the battery voltage of $k+i$ in the prediction domain is taken as ${{U}_{batt\_k}}$ of step $k$, similarly, the battery internal resistance is taken as ${{R}_{batt\_k}}$, and the fuel cell power is taken as ${{P}_{fc\_k}}$, which is as follows:
\begin{equation}\label{E:Equ30}
z(k|k)=\cdots =z(k+n-1|k)={{[{{U}_{batt\_k}},{{R}_{batt\_k}},{{P}_{fc\_k}}]}^{T}}
\end{equation}

To realize this concept, in $Data\_table$, the data table $Data\_table\_filter1$ is obtained by filtering the input sequence that satisfies the three characteristic values of ${{U}_{batt\_table}}={{U}_{batt\_k}}$, ${{R}_{batt\_table}}={{R}_{batt\_k}}$, and ${{P}_{fc\_table}}={{P}_{fc\_k}}$. Therefore, only the three-dimensional data table between the input of vehicle demand power ${{P}_{load\_table}}$ and the battery power ${{P}_{batt\_table}}$ and the output of $\Delta SOC$ is included in the $Data\_table\_filter1$. And then, the state observation equation can be expressed as follows:
\begin{equation}\label{E:Equ31}
\begin{aligned}
&\dot{x}=h(u,v,z) \\ 
&where\text{ }z={{[{{U}_{batt\_k}},{{R}_{batt\_k}},{{P}_{fc\_k}}]}^{T}} \\ 
\end{aligned}
\end{equation}

After obtaining the estimation sequence of vehicle power demand through vehicle velocity estimation, the data table $Data\_table\_filter1$ is further filtered to meet the vehicle demand power ${{P}_{load\_table}}$ in the data table is equal to the estimated variable of step $i$ in the estimation sequence, and the data table $Data\_table\_filter2$ is obtained. Furthermore, only the two-dimensional mapping relationship between the input of the battery power ${{P}_{batt\_table}}$ and the output of $\Delta SOC$ is included in the $Data\_table\_filter2$, which is as follows:
\begin{equation}\label{E:Equ32}
\begin{aligned}
  & \dot{x}=h(u,v,z) \\ 
 & where\text{ }\left\{ \begin{aligned}
  & v={{P}_{load}}(k|k),\cdots {{P}_{load}}(k+n-1|k) \\ 
 & z={{[{{U}_{batt\_k}},{{R}_{batt\_k}},{{P}_{fc\_k}}]}^{T}} \\ 
\end{aligned} \right. \\ 
\end{aligned}
\end{equation}

Next, the seventh-order equation is used to fit the above two-dimensional mapping relationship of $Data\_table\_filter2$ for further application in the online control of MPC. Therefore, the following equation can be obtained:
\begin{equation}\label{E:Equ33}
\begin{aligned}
  & \dot{x}=h(u,v,z)=\sum\limits_{j=0}^{7}{{{a}_{j}}{{u}^{j}}} \\ 
 & where\text{ }\left\{ \begin{aligned}
  & v={{P}_{load}}(k|k),\cdots,{{P}_{load}}(k+n-1|k) \\ 
 & z={{[{{U}_{batt\_k}},{{R}_{batt\_k}},{{P}_{fc\_k}}]}^{T}} \\ 
\end{aligned} \right.
\end{aligned}
\end{equation}
\begin{equation}\label{E:Equ34}
\begin{aligned}
  & x(k+i|k)=x(k+i-1|k)+T\cdot \sum\limits_{j=0}^{7}{{{a}_{j}}{{u}^{j}}(k+i-1|k)} \\ 
 & where\text{ }\left\{ \begin{aligned}
  & v(i)={{P}_{load}}(k+i-1|k) \\ 
 & z={{[{{U}_{batt\_k}},{{R}_{batt\_k}},{{P}_{fc\_k}}]}^{T}} \\ 
\end{aligned} \right.
\end{aligned}
\end{equation}

\begin{algorithm}[H]
\caption{Online control algorithm of LRMPC strategy}\label{A:Alg1}
\begin{algorithmic}[1]
\renewcommand{\algorithmicrequire}{\textbf{Input:}}
\REQUIRE ${{U}_{batt\_k}}$, ${{R}_{batt\_k}}$, ${{P}_{fc\_k}}$, $SO{{C}_{k}}$, ${{P}_{load}}$, $Data\_table$
\renewcommand{\algorithmicrequire}{\textbf{Output:}}
\REQUIRE ${{u}_{result}}\left( 1 \right)$
\renewcommand{\algorithmicrequire}{\textbf{Notations:}}
\REQUIRE ${{U}_{batt\_k}}$ = Battery voltage in step $k$\\
${{R}_{batt\_k}}$ = Battery resistance in step $k$\\
${{P}_{fc\_k}}$ = Fuel cell power in step $k$\\
$SO{{C}_{k}}$ = State of charge in step $k$\\
${{P}_{load}}$ = Vehicle demand power estimation sequence\\
$Data\_table$ = Data table from ML model\\
${{u}_{result}}\left( 1 \right)$ = Control results
\renewcommand{\algorithmicrequire}{\textbf{Initialize:}}
\REQUIRE ${{u}_{result}} = [], SOC\left( 0 \right) =~SO{{C}_{k}}$
\WHILE{${{P}_{load}} >$ 0}
    \STATE ${{u}_{0}} = 1-25/{{P}_{load}}$
    \STATE ${{u}_{min}} = {{u}_{0}}-{{k}_{1}}$
    \STATE ${{u}_{max}} = {{u}_{0}}+{{k}_{1}}$
    \STATE $\left[ M,N \right] = ~\left[ {{P}_{load}},{{P}_{fc\_max}}-{{P}_{load}} \right]$
    \STATE $Data\_table\_filter1$ = Filter $Data\_table$ satisfying ${{U}_{batt\_table}}={{U}_{batt\_k}}$, ${{R}_{batt\_table}}={{R}_{batt\_k}}$, and  ${{P}_{fc\_table}}={{P}_{fc\_k}}$.
    \STATE $i$ = 1
    \WHILE{$i \le n$}
        \STATE ${{P}_{pre}} = {{P}_{load}}\left( i \right)$
        \STATE $Data\_table\_filter2$ = Filter $Data\_table\_filter1$ satisfying ${{P}_{load\_table}}={{P}_{pre}}$.
        \STATE Fitting curves$~model\left( i \right)$ using ${{P}_{batt\_table}}$ and $\Delta SOC$.
        \STATE Update the state $SOC\left( i \right)$ using $model\left( i \right)$ to the function of $SOC\left( i-1 \right)+\text{ }\Delta SOC\left( {{P}_{batt}}\left( i \right) \right)\cdot step\_time$.
    \ENDWHILE
    \STATE ${{u}_{result}}$ = argmin($J$) subject to$~\left[ {{u}_{min}},{{u}_{max}} \right]$ and $Mu$+$N$
    \STATE $i$ = $i+1$
\ENDWHILE
\RETURN ${{u}_{result}}\left( 1 \right)$
\end{algorithmic}
\end{algorithm}

Same as the traditional MPC, LRMPC also acquires the optimal control trajectory through the solution process in Eq. \eqref{E:Equ25}. Explicit data tables obtained from offline training can be applied in online control, not only can accurately reflect the real response characteristics of the state observation model to achieve accurate control and energy saving but also can greatly reduce the solution time, with high potential for online application.

\subsection{Reference velocity estimation based on deep forest}
The deep forest is a decision tree ensemble approach proposed in 2017, which is also known as the multi-grained cascade forest. The deep forest has the following advantages: a) the training process is efficient and scalable; b) there are few hyperparameters and the model is insensitive to hyperparameter adjustment; c) the model complexity can be adaptively scaled according to the size of the dataset; d) each cascade is generated using cross-validation to avoid overfitting. 

The above advantages are due to the following key steps in regression analysis of the deep forest: the multi-grained scanning procedure (MGSP), the cascade forest procedure (CFP), and the cross-validation procedure (CVP).

\begin{figure*}[!t]
\centering
\includegraphics[width=5.7in]{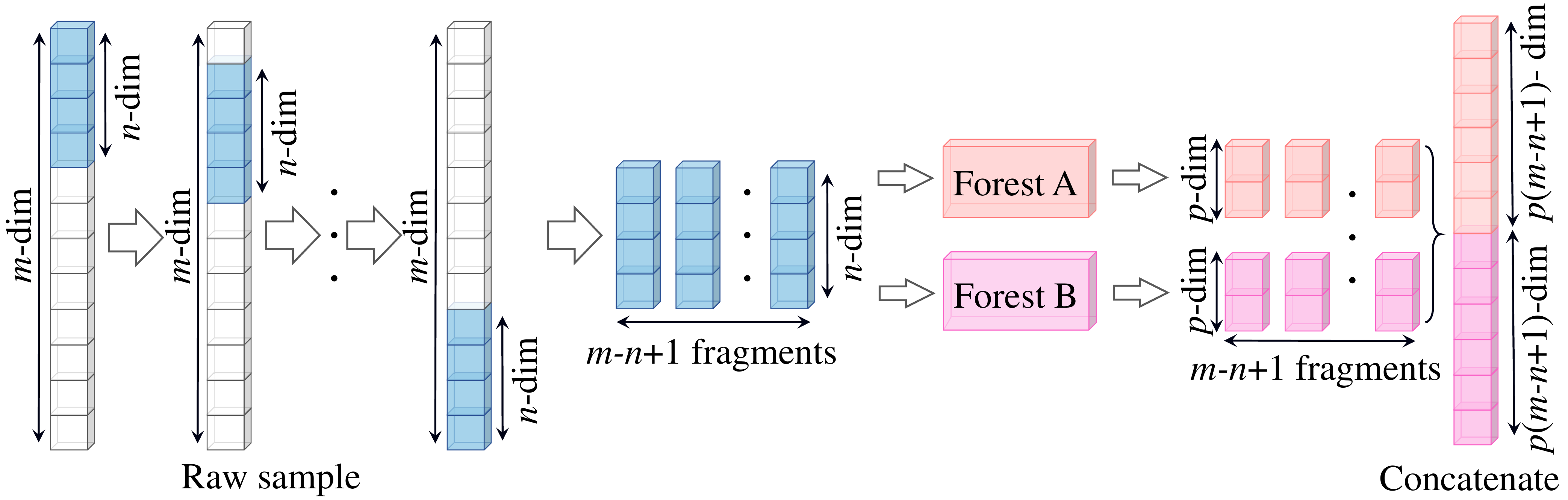}
\caption{Procedure of multi-grained scanning.}
\label{F:Fig6}
\end{figure*}

MGSP can extract the features of samples and mine the feature vectors of sequence data. As shown in Fig. \ref{F:Fig6}, assuming that the dimension of the feature vector without MGSP processing is $m$, $m-n+1$ sets of feature data can be obtained with a scanning window of n dimensions, and these $m-n+1$ sets of n-dimensional feature data are processed by complete-random tree forest and random forest, and finally, a total of $2p\left( m-n+1 \right)$-dimensional feature vectors are obtained as the input of CFP. MGSP can enhance cascading forests, and the enhancement becomes more pronounced as the scan window changes from single to multiple sizes.

\begin{figure*}[!t]
\centering
\includegraphics[width=5.7in]{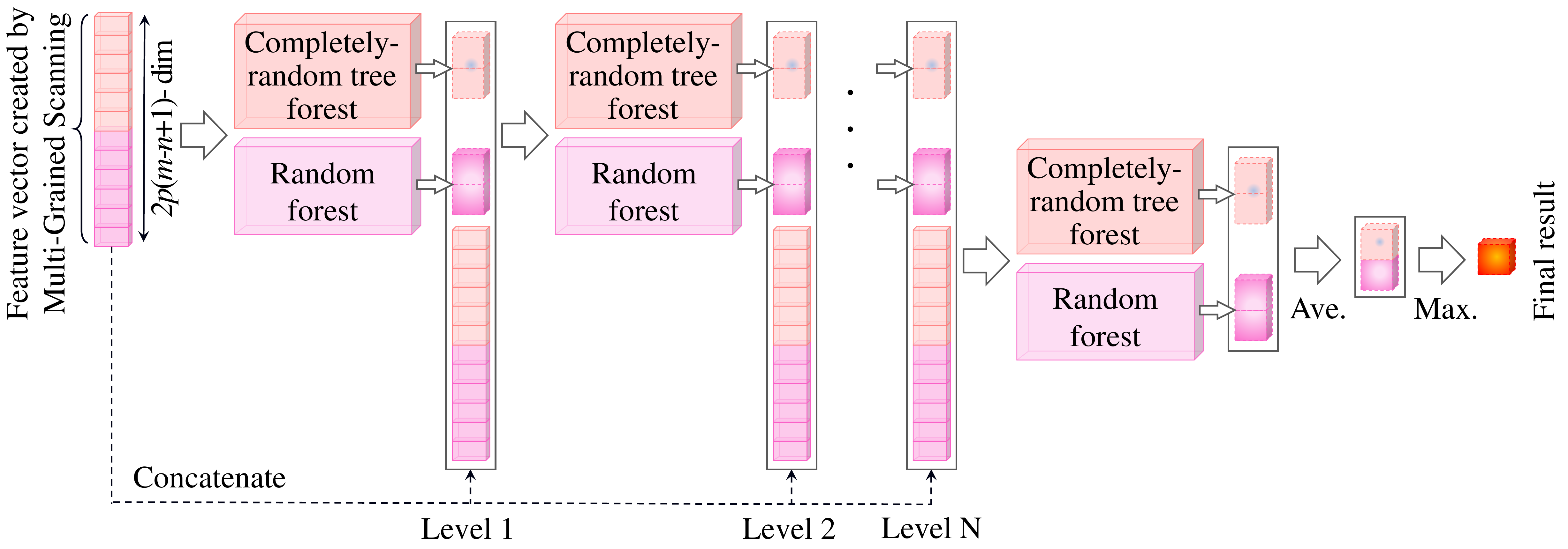}
\caption{Procedure of cascade forest.}
\label{F:Fig7}
\end{figure*}

CFP can improve the accuracy of prediction. To ensure and improve the generalization of the prediction model, the cascade forest consists of a combination of several layers including different types of decision tree forest models. Each layer in this study consists of two complete-random tree forests with default parameters and two random forests, as shown in Fig. \ref{F:Fig7}. The first layer directly employs the feature vectors obtained from MGSP as training inputs, and the output two-dimensional vectors of both types are called augmented feature vectors due to their effective representation learning of the sample features in the deep forest. The training inputs of the subsequent layers are concatenated from the augmented feature vectors of the previous layer and the feature vectors obtained from MGSP until the model results of the final layer are obtained.

To mitigate the overfitting of the CFP in deep forest regression, k-fold cross-validation is applied to validate the vectors generated by each forest of the CFP. After extending a new layer, if there is no significant performance improvement in the evaluation of the whole cascade performance, it means that the nodes have reached full purity, at which point the training process will be terminated and the final depth of the obtained deep forest model is equal to the number of cascade forests. Therefore, compared with DNN, the number of layers of the deep forest model is automatically determined by the training process, which can alleviate the complexity of the model and the sensitivity of hyperparameter tuning.

In generating the reference trajectory, the input features of the deep forest model are the vehicle velocity and acceleration at moment $k$, and the output is the vehicle velocity and acceleration at moment $k+1$. And then, the estimated velocity sequence is processed into the demand power sequence for MPC state observation during the online application. The process is as follows:
\begin{equation}\label{E:Equ35}
{{F}_{t\_pre}}(k|k)=mgf+\frac{1}{2}{{C}_{D}}AV_{t\_pre}^{2}(k|k)+\delta m\frac{d{{V}_{t}}}{dt}
\end{equation}
\begin{equation}\label{E:Equ36}
{{P}_{t\_pre}}(k|k)={{F}_{t}}(k|k)\cdot {{V}_{t}}(k|k)
\end{equation}
\begin{equation}\label{E:Equ37}
{{P}_{load\_pre}}(k|k)=\frac{{{P}_{t\_pre}}(k|k)}{{{\eta }_{DC/AC}}{{\eta }_{motor}}}
\end{equation}
where ${{F}_{t\_pre}}$ is the vehicle drive force estimation sequence, ${{P}_{t\_pre}}$ is the vehicle drive power estimation sequence, and ${{P}_{load\_pre}}$ is the vehicle demand power estimation sequence.

\section{Simulation and discussion}
To validate the comprehensive performance of the proposed LRMPC, a series of simulations are performed based on MATLAB R2021b and Python 3.8, including the verification of vehicle velocity estimation and energy-saving ability and real-time application of LRMPC. In the verification of vehicle velocity estimation, all training datasets used in the vehicle velocity estimation models are made up of driving cycles collected from the actual traffic scenes. In the verification of LRMPC, the comprehensive performance of LRMPC is verified by comparing the energy-saving ability and real-time application capability under different lengths of prediction domains with the benchmark method of traditional MPC (T-MPC) \cite{ref43}. As a result, the capability of the LRMPC strategy for the 4WD FCEV is comprehensively evaluated. Note that the simulations are performed on a computer equipped with an Intel i5-6300HQ processor and an 8 GB memory.
\subsection{Performance verification of vehicle velocity estimation based on deep forest}
Since the future vehicle velocity directly impacts on the reference obtaining, which could further affect the energy-saving performance of energy management strategies in the 4WD FCEV, analyzing the accuracy of vehicle velocity estimation is necessary. The performance of the vehicle velocity estimation is evaluated via comparing with some baseline velocity estimations, including:
\begin{itemize}
\item{ARIMA: Autoregressive integrated moving average model (ARIMA) is a time series prediction and analysis method, which can convert non-stationary time series into stationary time series through differential processing \cite{ref44}. The autoregressive coefficient is set to 3, the moving average coefficient is set to 4, and the differential order is set to the second order.}
\item{LSSVM: The improved algorithm based on the standard support vector machine (SVM) transforms the convex quadratic programming problem into a linear problem in the training process and takes the maximum-margin hyperplane as the decision boundary, which can accelerate the calculating speed \cite{ref45}. The regularization parameter of the radial basis kernel is set to 10.}
\item{GPR: A non-parametric model for regression analysis using the gaussian process (GP) \cite{ref46}. GPR is theoretically a universal approximator of any continuous function in compact space, and Bayesian inference is used in the solving process. The fitting and prediction method are set to exact gaussian process regression, and the square exponential kernel is used.}
\end{itemize}

As for the deep forest model, it should be specially noted that the number of per decision trees is set to 100 and the maximum number of layers is set to 20.

TABLE \ref{T:Table2} shows the numerical comparison results between different prediction models with different estimation lengths, where MAE and RMSE represent mean absolute error and root mean square error, respectively.

\begin{table}[!t]
\caption{Comparison of different methods \\with different estimation lengths\label{T:Table2}}
\centering
\setlength{\tabcolsep}{1mm}{
\begin{tabular}{c c c c}
\toprule
Methods                      & Estimation lengths & MAE    & RMSE   \\ \hline
                             & 1s                 & 0.1223 & 0.2231 \\
\multirow{3}{*}{Deep forest} & 2s                 & 0.4504 & 0.8677 \\
                             & 3s                 & 0.8995 & 1.5380 \\
                             & 1s                 & 0.3029 & 0.5718 \\
\multirow{3}{*}{ARIMA}       & 2s                 & 0.7882 & 1.3442 \\
                             & 3s                 & 1.4253 & 2.3021 \\
                             & 1s                 & 0.8705 & 1.3496 \\
\multirow{3}{*}{LSSVM}       & 2s                 & 1.7042 & 2.6252 \\
                             & 3s                 & 2.5067 & 3.8257 \\
                             & 1s                 & 0.8508 & 1.3469 \\
\multirow{3}{*}{GPR}         & 2s                 & 1.6830 & 2.6184 \\
                             & 3s                 & 2.4959 & 3.8154 \\
\bottomrule 
\end{tabular}}
\end{table}

To compare the general estimation performance of various velocity estimations, TABLE \ref{T:Table2} is provided to support analysis. As shown in TABLE \ref{T:Table2}, deep forest has the lowest estimation error compared with other estimation methods, which reaches 0.2231, 0.8677 and 1.5380 in 1s, 2s and 3s, respectively. This phenomenon shows the suitableness of deep forest in vehicle velocity estimation. LSSVM and GPR show the similar performance in velocity estimation, whose predictive errors are larger than other methods. The errors in these two methods number nearly 3 times that of the ones in deep forest. In addition, ARIMA has the moderate estimation accuracy, whose errors number around twice as much as the errors in deep forest.

\begin{figure}[!t]
\centering
\includegraphics[width=3.6in]{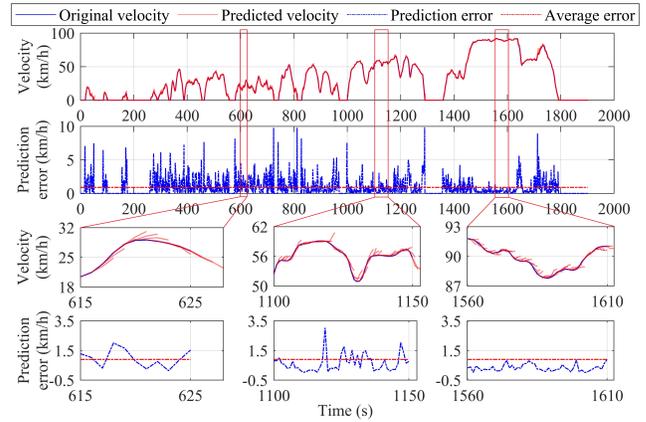}
\caption{Velocity estimation based on deep forest.}
\label{F:Fig8}
\end{figure}

To intuitively analyze the performance of the deep forest for velocity estimation, the performance with the estimation length of 3s is focused. The estimation curves and estimation errors of deep forest in 3s are illustrated in Fig. \ref{F:Fig8}, which has some enlarged graphs to provide detail information. 

Firstly, the enlarged graph on the left shows the details of the velocity estimation from 615s to 625s. The estimated velocity shows a more desirable prediction trend in this period as the predictive velocity curves are all near the natural velocity with the similar tendency, resulting in that the predictive error fluctuates in a smaller range from the baseline of the mean error. 

Secondly, the estimation results of the rapid velocity changing are shown in the middle-enlarged graph from 1100s to 1150s. The original velocity can be abstracted into the following five states: acceleration, uniformity, deceleration, acceleration, and uniformity. The velocity estimation error enlarges violently near the abrupt change point between two different states, especially when the velocity transforming from uniformity to deceleration and the velocity from deceleration to stabilize. 

Thirdly, the enlarged graph on the right side shows the performance of velocity estimation with little variation of the original velocity, and it can be found that the stable velocity brings higher prediction accuracy, leading to the small gap between the mean error line and prediction error curve. 

As a result, the estimated vehicle velocity obtained by the deep forest can meet the error requirements of online applications of MPC.

\subsection{Evaluation of LRMPC}
\begin{figure}[!t]
\centering
\includegraphics[width=3.4in]{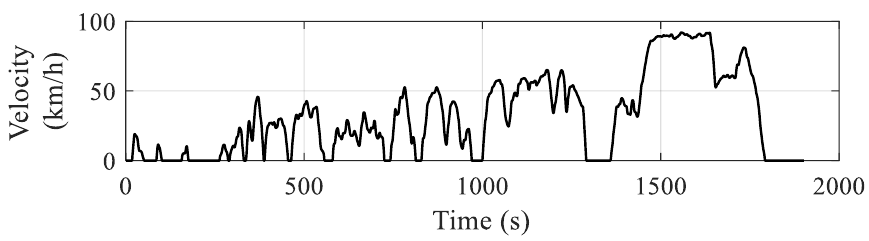}
\caption{Velocity-time curve of driving cycle.}
\label{F:Fig9}
\end{figure}

After analyzing the performance of vehicle velocity estimation, the control effect of LRMPC in 4WD FCEV can be further analyzed more efficiently. To evaluate the comprehensive performance of LRMPC, the analyzing aspects should include general comparison among the strategies with various machine learning methods generating explicit data tables, the state changing of components in LRMPC and the real-time practical capacity of LRMPC. Note that all simulation processes are working under the CLTC-C driving cycle (Fig. \ref{F:Fig9}) and all mentioned methods have a fixed simulation step of 0.05s.

The various machine learning methods used for constructing LRMPC include:
\begin{itemize}
\item{RF: A regression model integrated based on the classification and regression trees (CARTs), which is competent for regression tasks \cite{ref47}. The number of decision trees and leaf nodes are respectively set to 100 and 2.}
\item{NN: The back propagation (BP) neural network based on multilayer feed-forward \cite{ref48} is composed of input layer, hidden layer and output layer. And the BP training is conducted based on the gradient descent method to minimize the mean square error. The number of input layer nodes, hidden layer nodes and output layer nodes are respectively set to 5, 12, 1, and the learning rate and training accuracy requirement are respectively set to $1\times {{10}^{-10}}$ and $1\times {{10}^{-5}}$.}
\item{XGBoost: An improved algorithm of the gradient boosting \cite{ref49}. The loss function added with the regularization term is expanded to the second order by Taylor and the extreme value is solved by Newton method. The number of regression trees based on the linear model is set to 500 and the maximum depth is set to 10.}
\end{itemize}

\begin{figure}[!t]
\centering
\includegraphics[width=3in]{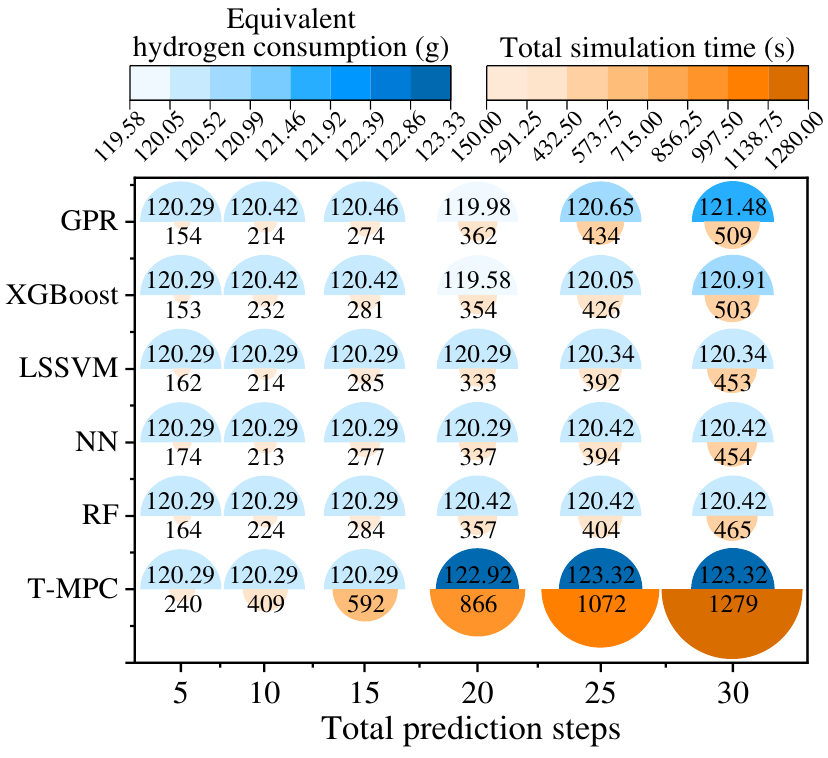}
\caption{Comparison results of different predicted steps.}
\label{F:Fig10}
\end{figure}

{\bf General evaluation of LRMPC: }To analyze the energy-saving performance of LRMPC and the suitableness of various machine learning methods used for generating explicit data tables, the simulations of LRMPC based on 5 machine learning methods and 6 types of predictive domains are exploited. The results presented in Fig. \ref{F:Fig10} and TABLE \ref{T:Table3} support the conclusion that the LRMPC controller is superior to the normal controller. Fig. \ref{F:Fig10} shows the comparison results for increasing the prediction domains from 5 steps to 30 steps, including the equivalent hydrogen consumption characterizing the energy savings, and the total simulation time characterizing the real-time performance. 

\begin{table}[!t]
\caption{Optimality of different prediction steps\label{T:Table3}}
\centering
\setlength{\tabcolsep}{1mm}{
\begin{tabular}{cccccccc}
\toprule
\multicolumn{8}{c}{Optimality (\%)}                                                      \\ \hline
\multicolumn{2}{c}{\multirow{2}{*}{Method}} & \multicolumn{6}{c}{Total prediction steps} \\
\multicolumn{2}{c}{}                        & 5  & 10    & 15    & 20    & 25    & 30    \\ \hline
\multicolumn{2}{c}{T-MPC}                   & 0  & 0.11  & 0.14  & 0     & 0     & 0     \\ \hline
\multirow{5}{*}{LRMPC}       & RF           & 0  & 0.11  & 0.14  & 2.03  & 2.35  & 2.35  \\
                             & NN           & 0  & 0.11  & 0.14  & 2.14  & 2.35  & 2.35  \\
                             & LSSVM        & 0  & 0.11  & 0.14  & 2.14  & 2.42  & 2.42  \\
                             & XGBoost      & 0  & 0     & 0.03  & 2.72  & 2.65  & 1.95  \\
                             & GPR          & 0  & 0     & 0     & 2.39  & 2.17  & 1.49  \\
\bottomrule 
\end{tabular}}
\end{table}

It is obvious that there is little difference in the equivalent hydrogen consumption between T-MPC and the other five LRMPC control methods when the number of prediction domains is between 5 steps and 15 steps. Whereas, the energy-saving advantage of LRMPC is obvious after the number of prediction steps is increased to 20 steps with general energy-saving optimality exceeding 2$\%$. The reason is that LRMPC can completely reduce energy consumption by obtaining the accurate optimal solution in the prediction domains, which is caused by an accurate prediction of the future driving state in the specified range. 

By analyzing the total simulation time, it is observed that in the process of changing the prediction domains from 5 steps to 30 steps, the longer the length of prediction domains, the higher the computational cost. Besides, the advantages of LRMPC over T-MPC are more obvious with the length of prediction domains increasing. Specifically, the total simulation time of T-MPC and LRMPC even increases by nearly 5 times and 3 times respectively. In addition, when the prediction steps are the same, for example, when the prediction domain is 20 steps, the total simulation time of the five LRMPC methods is basically the same, reaching about 350s, which is shorter than the 866s of T-MPC. This is because all LRMPC methods obtain the higher-order mapping relationships by looking up the explicit data tables of the same dimension, and such a similar element combination can provide the solution convenience for the accumulation of state changes in the cost function. In contrast, the cost function of T-MPC is obtained by the linearized equation containing the root equation, which can create trouble for the combined solution of the state changes at each step, thus lengthening the total simulation time.

{\bf State changing of components in LRMPC: }To better prove the energy-saving advantages brought by the component state of LRMPC, the simulation process with the prediction domain of 20 steps is especially discussed including curves of component power, component hydrogen consumption, battery SOC and internal resistance and simulation results of state trajectories and control trajectories. As shown in Figs. 11 to 15.

\begin{figure}[!t]
\centering
\includegraphics[width=3.4in]{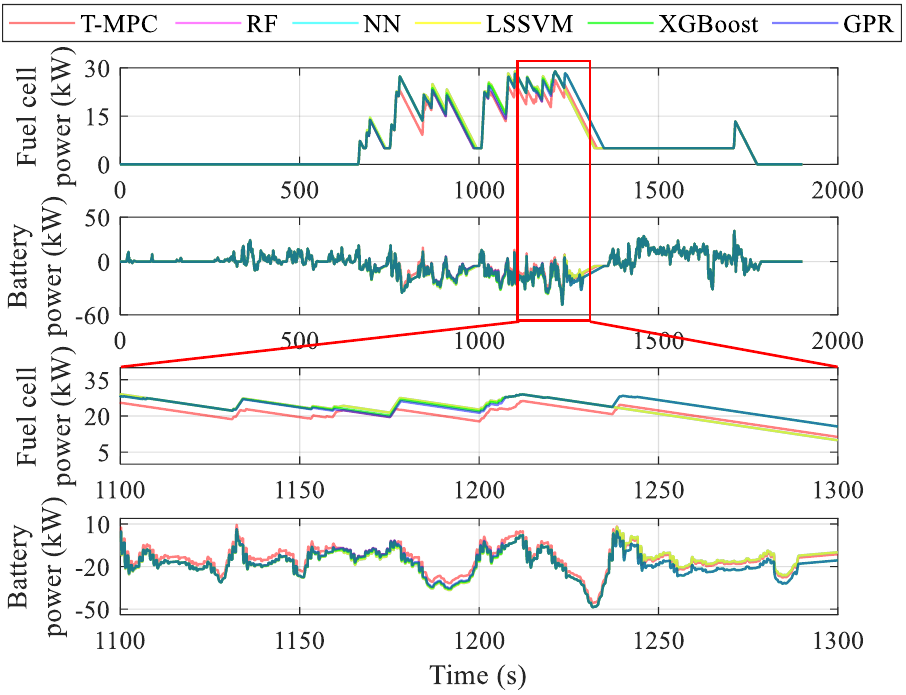}
\caption{Fuel cell power and battery power trajectories by different methods.}
\label{F:Fig11}
\end{figure}

Fig. \ref{F:Fig11} shows power curves of fuel cell and battery concerning time, and it is observed that the power of fuel cell and battery are similar among the different methods in EV mode before about 750s. Further, the graphs in the range from 1100s to 1300s are scaled up to represent the differences more clearly in HEV mode, and the fuel cell power of LRMPC control methods is generally higher than those of the T-MPC with a maximum gap reaching around 7kW. From 1100s to 1300s, LRMPC integrating XGBoost explicit table data has the highest fuel cell power, resulting in more charging power for battery.

\begin{figure}[!t]
\centering
\includegraphics[width=3.4in]{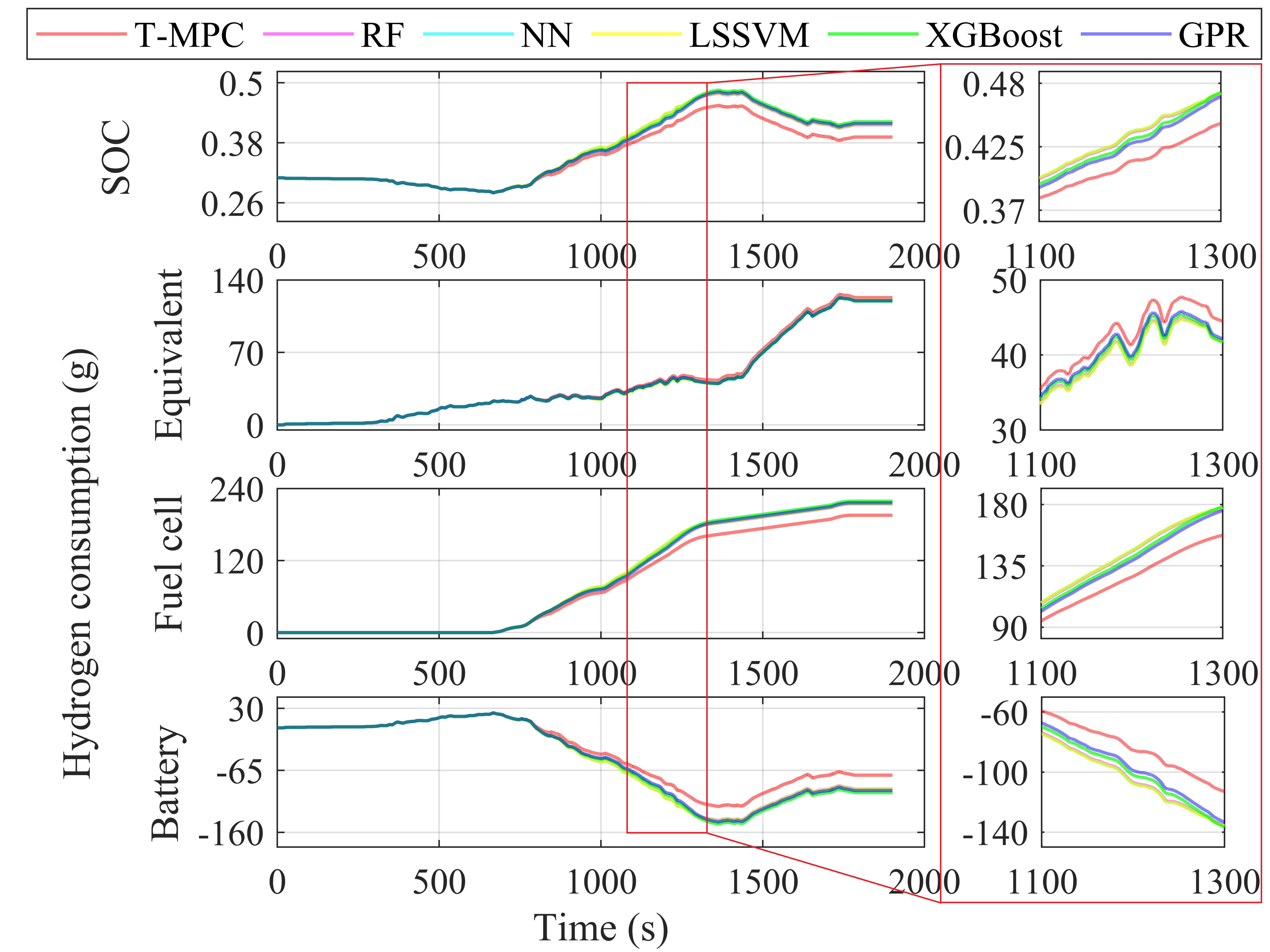}
\caption{Battery SOC and fuel consumption trajectories by different methods.}
\label{F:Fig12}
\end{figure}

Fig. \ref{F:Fig12} shows the curves of battery SOC, equivalent hydrogen consumption, fuel cell hydrogen consumption, and battery hydrogen consumption to visually compare the final energy consumption results. Same as above, all control methods have the same control results before the vehicle enters HEV mode. As for the simulation results in HEV mode from 1100s to 1300s, the change of battery SOC obtained by T-MPC is the smallest, increasing about 0.062. This phenomenon results in that the change of final equivalent hydrogen consumption is the largest, increasing by about 11.07g. Conversely, the LRMPC integrating XGBoost has the largest SOC change, increasing 0.079 from 1100s to 1300s. Thus, the change of equivalent hydrogen consumption is the smallest with 7.65g. In addition, the gap in SOC trajectories between T-MPC and the LRMPC integrating XGBoost expands with time, which changed from the difference of 0.011 in 1100s to 0.028 in 1300s. Similarly, the gap in equivalent hydrogen consumption trajectories between T-MPC and the LRMPC integrating XGBoost changes from 0.63g in 1100s to -2.79g in 1300s. Two aspects directly affecting the equivalent hydrogen consumption results are further analyzed, including fuel cell hydrogen consumption and equivalent hydrogen consumption for battery power. It can be found that even if the higher fuel cell power of LRMPC brings more fuel cell hydrogen consumption, the higher battery charging power leads to a larger equivalent hydrogen consumption for battery power. Thus, the offsetting effect of the equivalent hydrogen consumption results is enlarged. 

\begin{figure}[!t]
\centering
\includegraphics[width=3.4in]{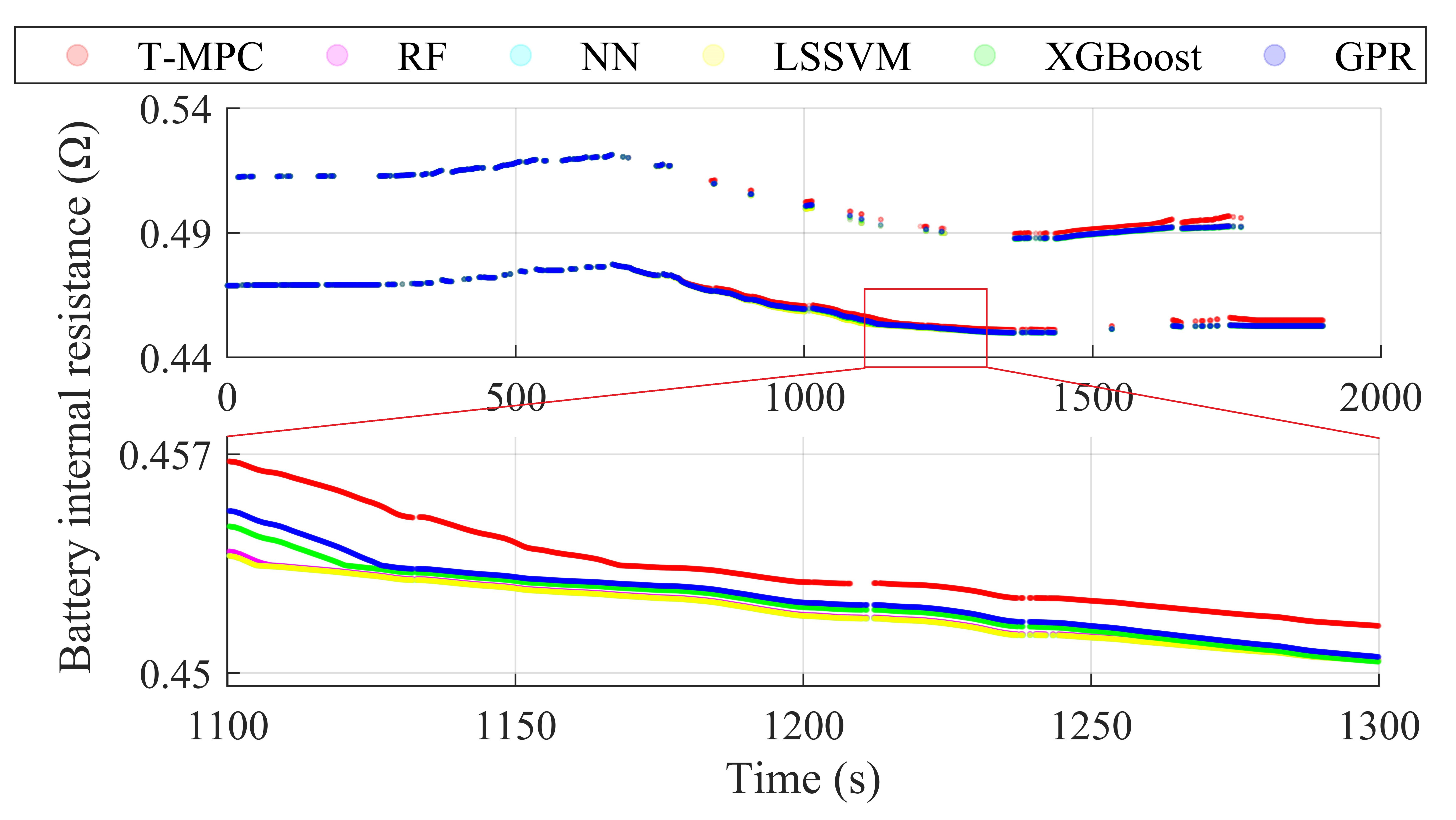}
\caption{Battery internal resistance trajectories by different methods.}
\label{F:Fig13}
\end{figure}

Fig. \ref{F:Fig13} shows the internal resistance distribution of battery. A series of scatter points distributed between 0.49$\Omega$ and 0.54$\Omega$ is the battery discharging internal resistance trajectory and the other one distributed between 0.44$\Omega$ and 0.49$\Omega$ is the battery charging internal resistance trajectory. From the magnified graph in Fig. \ref{F:Fig13}, thanks to the input information about the internal resistance of the battery in LRMPC explicit data tables, all LRMPC methods ensure that the battery operates at a lower internal resistance during the control process. To be specific, the maximum internal resistance difference between T-MPC and LRMPC is about 0.003$\Omega$. Therefore, the smaller internal resistance of charge is supposed to bring more significant energy-saving advantages in LRMPC. As a result, the control result of better energy consumption is closely related to the battery operating state.

\begin{figure}[!t]
\centering
\includegraphics[width=3.4in]{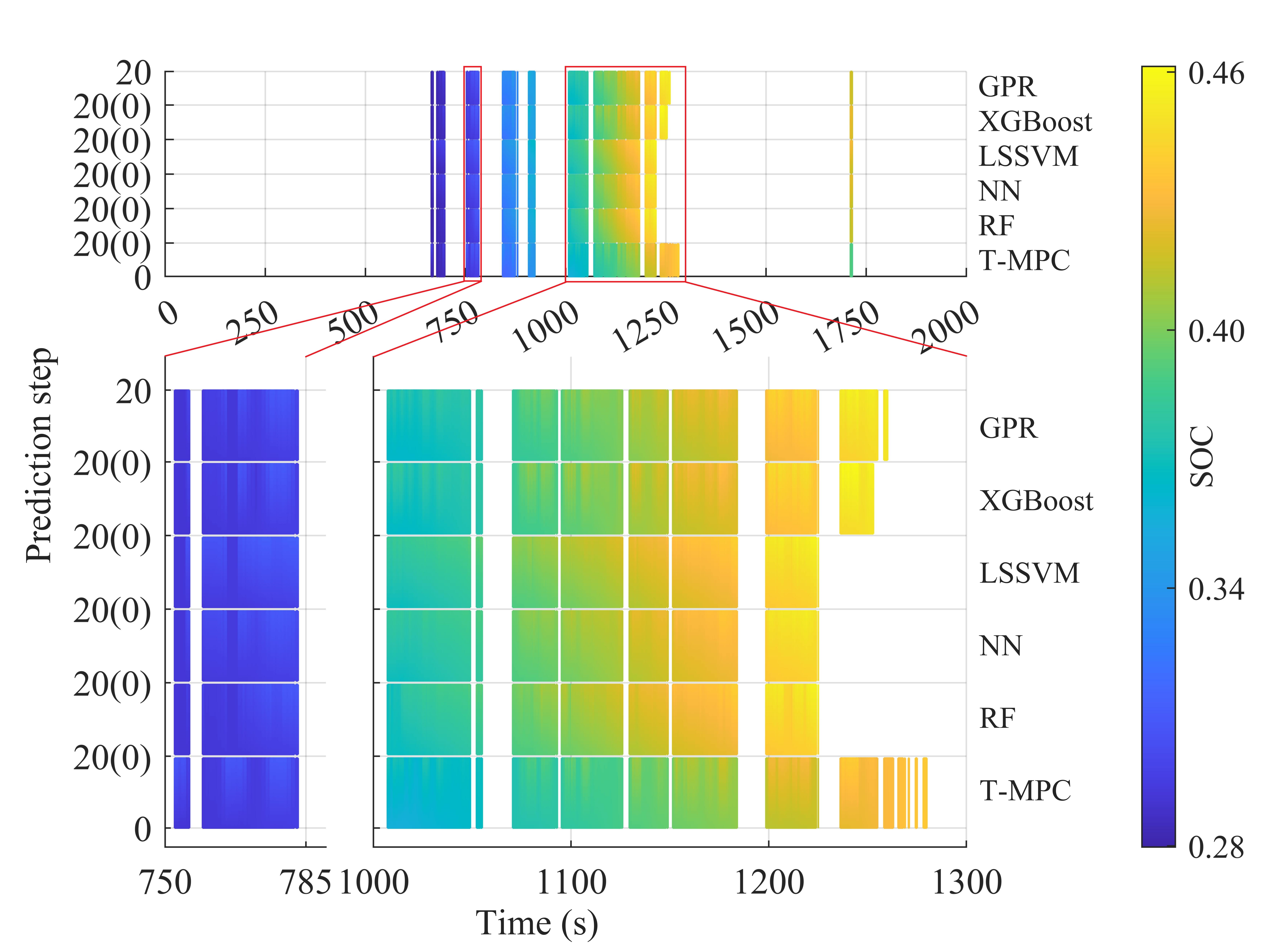}
\caption{State prediction trajectories by different methods.}
\label{F:Fig14}
\end{figure}

\begin{figure}[!t]
\centering
\includegraphics[width=3.4in]{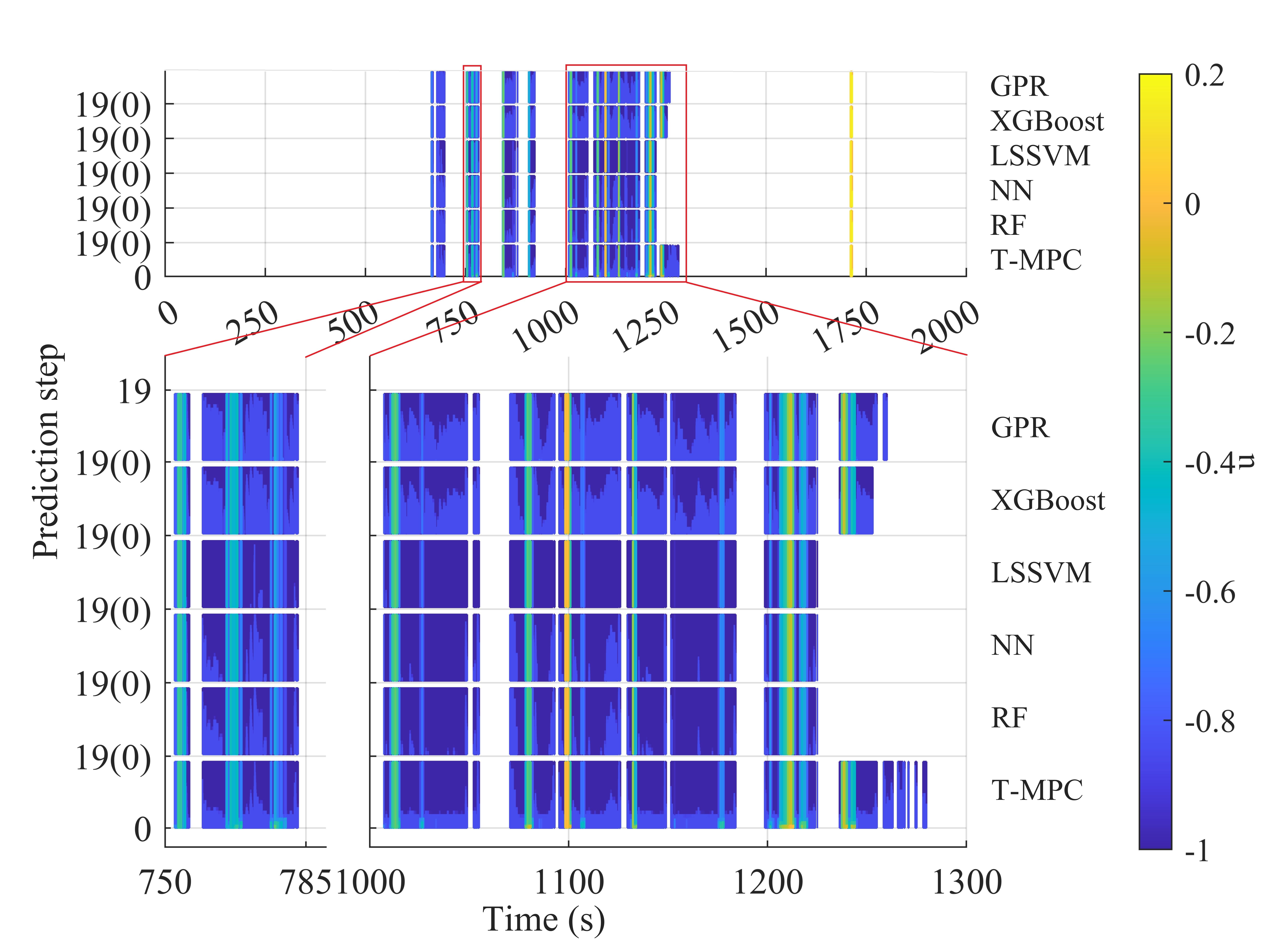}
\caption{Control trajectories by different methods.}
\label{F:Fig15}
\end{figure}

Moreover, more accurate state observation can lead to better control trajectories, and better energy consumption control results can be obtained through the interaction of the two. Fig. \ref{F:Fig14} and Fig. \ref{F:Fig15} show the state trajectories and control trajectories of different control methods in the prediction domains under HEV mode, respectively. The single-step state transfer of LRMPC is obviously larger than that of T-MPC under the same time and similar control trajectory, and the results of T-MPC, RF, and NN at the 1100s are enough to prove this argument. In addition, by comparing the results of XGBoost or GPR with LRMPCs embedding other machine learning methods, the control trajectories of XGBoost and GPR have a greater impact on state transfer. This can be explained that XGBoost and GPR can achieve results as similar as the ones in other LRMPCs without applying a large amount of control. The explicit data tables established by XGBoost and GPR in this study are indirectly proven to be more accurate in reflecting the state observation model. Moreover, T-MPC has the longest controlling time, reaching 1280s in Fig. \ref{F:Fig15}. LRMPCs based on RF, NN, and LSSVM have the same controlling time, which is the shortest one numbered 1230s. This also indicates that aiming at achieving a relatively better performance of energy loss, T-MPC still needs to increase the charging process of battery in HEV mode.

\begin{figure}[!t]
\centering
\includegraphics[width=3.4in]{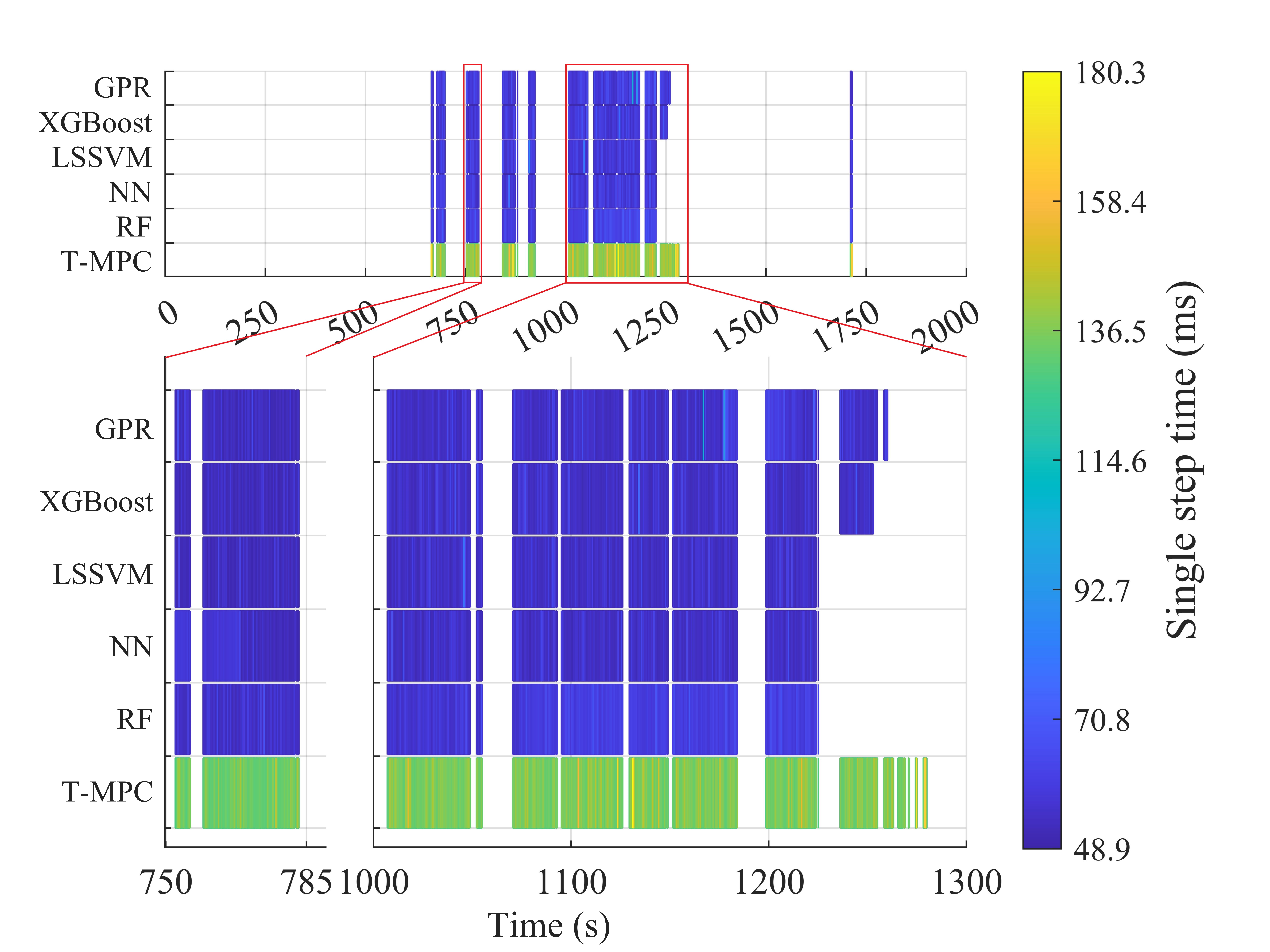}
\caption{Single-step time by different methods.}
\label{F:Fig16}
\end{figure}

{\bf Real-time capacity of LRMPC: }The real-time advantage of LRMPC is further verified by a comparative analysis of single-step simulation times in the prediction domain of 20 steps, as shown in Fig. \ref{F:Fig16}. As mentioned earlier, since the five LRMPC control methods perform state updates by using explicit data tables of the same dimension, the lengths of single-step simulation time in LRMPC with different machine learning methods are not significantly different and are all much smaller than the one of T-MPC, generally reaching half of that in T-MPC. The main reason is that the same form of higher-order mapping relationship of LRMPC used for state updating provides the possibility for the similar elements combination of cost functions. The comparison of single-step times reveals that LRMPC has better potential for real-time applications.

By comparing from different perspectives, the LRMPC controller proposed in this study shows excellent performance in FCEV energy management, with extraordinary control effects and real-time application potential.

\section{Discussion}
The above section validates the performance of vehicle velocity estimation based on deep forest and the optimal control performance of LRMPC. In order to further analyze the potential impact of LRMPC explicit solutions on real-time applications, several discussions are given as follows:

\begin{enumerate}
\item{Traditional MPC uses linearization methods to reflect the response characteristics of the control system. When the coefficient matrix of the linearized equation is fixed, it can be solved by quadratic programming and brings considerable real-time application performance, but the accuracy of state observation is sacrificed. However, when the coefficient matrix of the linearized equation is time-varying, the state observation accuracy is expected to be improved, but the solution efficiency is extremely low. Similarly, LRMPC faces almost the same contradiction between state observation accuracy and real-time performance.}
\item{Thanks to the strong generalization performance of machine learning models, it is beneficial to create more appreciable accuracy of state observation and more scalable control strategies by integrating machine learning models in MPC directly. However, the additional machine learning models inevitably reduce the solution efficiency by adding operations. Therefore, obtaining explicit solutions by training machine learning models offline can make it possible to further improve the performance of real-time applications for learning-based MPC.}
\item{In this study, the explicit solution in the LRMPC method is implemented as follows: a) the mapping relationship between the five-dimensional input and the one-dimensional output of the control system is obtained by training machine learning models offline; b) a specific range of discrete inputs defines the outputs of explicit data tables by machine learning models; c) two-dimensional higher-order fitting equations for state observation are obtained under the assumption of partial constant input characteristics in the prediction domains. Therefore, the multidimensional mapping relationship is sufficient to fully reflect the dynamic performance of the control system to bring about better energy-saving control results above.}
\item{As for real-time application performance, although this explicit solution has not yet reached the full explicit LRMPC, the unified form of the state components in the cost function of LRMPC is sufficient to significantly improve the real-time performance compared to the traditional MPC, which is consistent with the above comparison results. More efforts will be devoted to the explicit solution of LRMPC in our future work to exploit the potential of online applications.}
\end{enumerate}

\section{Conclusions}
In this paper, a novel LRMPC EMS is proposed for 4WD FCEVs to improve fuel economy and real-time application ability. The mapping relationship of the control system are constructed offline by five basic ML methods including RF, NN, LSSVM, XGBoost, and GPR. And then the explicit data tables characterizing the control laws are obtained and applied to the online control process of MPC to mitigate the adverse effects on state observation. In addition, the reference trajectory of vehicle velocity is estimated by the deep forest with a much smaller error in order to improve the control effect. This specific velocity estimation method can reasonably analyze future driving behavior without significant computational cost. Besides, the performance of LRMPC is verified in the prediction domains from 5 to 30 steps using the T-MPC as the benchmark strategy. The simulation results show that as the prediction domains increases, the superiority of LRMPC is gradually and synchronously demonstrated, verifying the robustness of LRMPC. Meanwhile, thanks to more accurate state observation, the energy-saving potential  is especially obvious in the prediction domain of 20 steps, where the optimal LRMPC strategy improves fuel economy by 2.72$\%$ and shortens the total running time by 61.55$\%$ compared with the baseline strategy of T-MPC. As a result, a better EMS for nonlinear complex systems is achieved.

\bibliographystyle{IEEEtran}
\bibliography{ref}

\newpage
\begin{IEEEbiography}[{\includegraphics[width=1in,height=1.25in,clip,keepaspectratio]{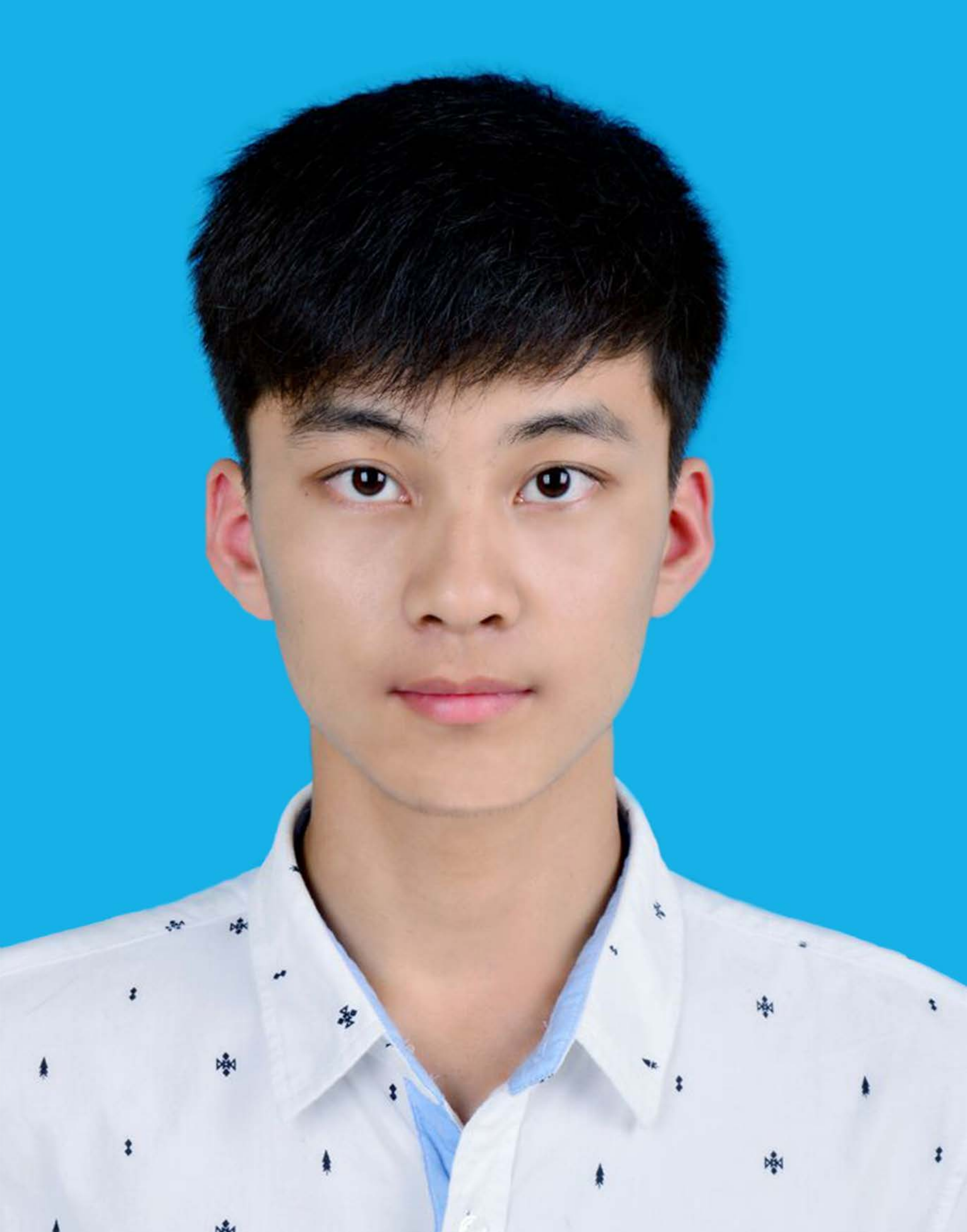}}]{Shibo Li} was born in Jiamusi City, Heilongjiang Province on May 8th, 1998. He received the B.E. degree in vehicle engineering from Jilin University, Changchun, China, in 2020. He is currently pursuing continuous academic program for a Ph.D. degree in vehicle engineering with Jilin University, Changchun, China. 

His research interests include basic machine learning, optimal energy management strategy about fuel cell electric vehicles.
\end{IEEEbiography}
\vspace{-10mm}

\begin{IEEEbiography}[{\includegraphics[width=1in,height=1.25in,clip,keepaspectratio]{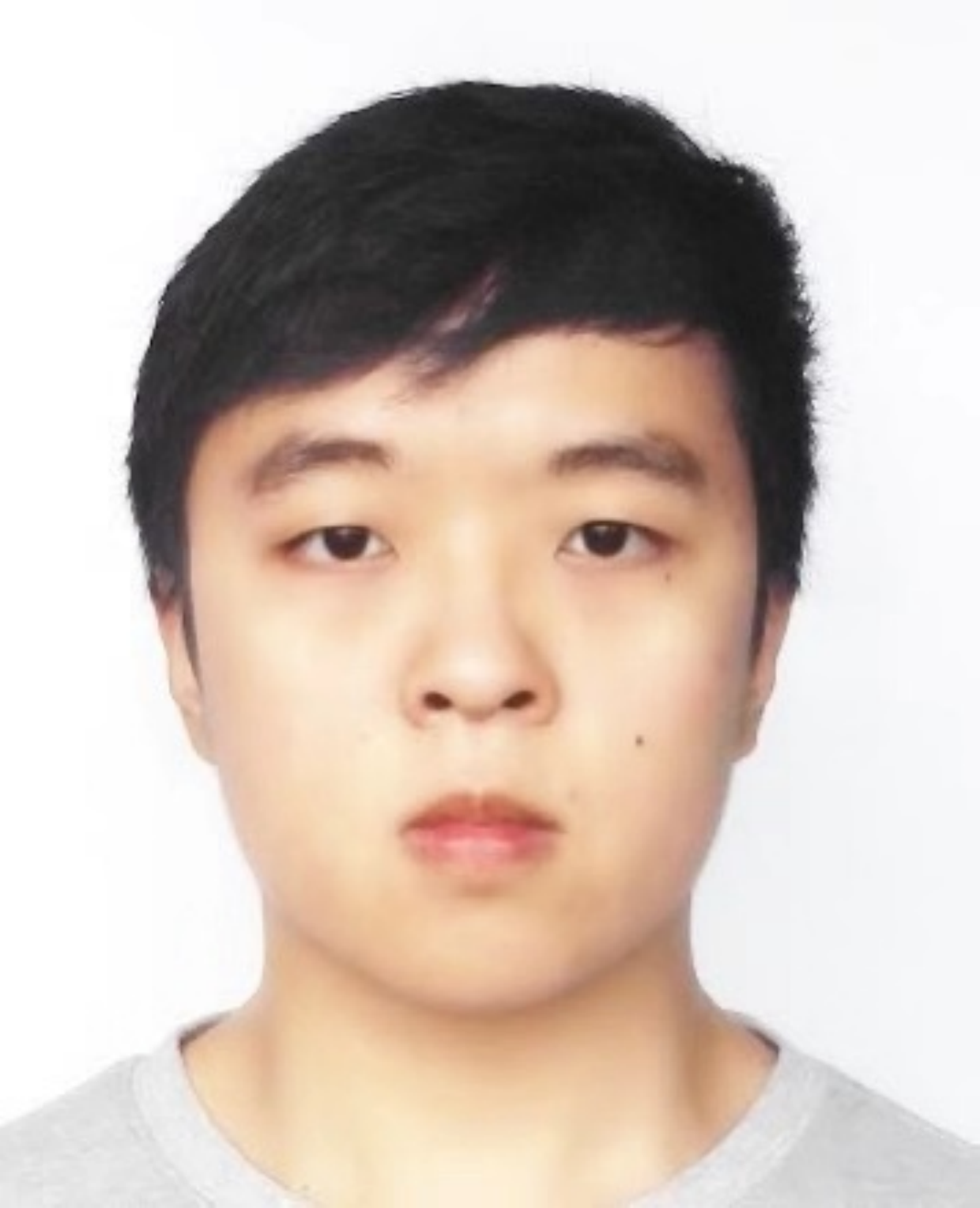}}]{Zhuoran Hou} received the B.S. degree in vehicle engineering from Chongqing University, Chongqing, China, in 2017. and the M.S. in Automotive Engineering from Jilin University, China, in 2020. He is currently pursuing continuous academic program involving doctoral studies in automotive engineering with Jilin University, Changchun, China. 

His research interests include basic machine learning, optimal energy management strategy about plug-in hybrid vehicles. 
\end{IEEEbiography}
\enlargethispage{-9cm}

\begin{IEEEbiography}[{\includegraphics[width=1in,height=1.25in,clip,keepaspectratio]{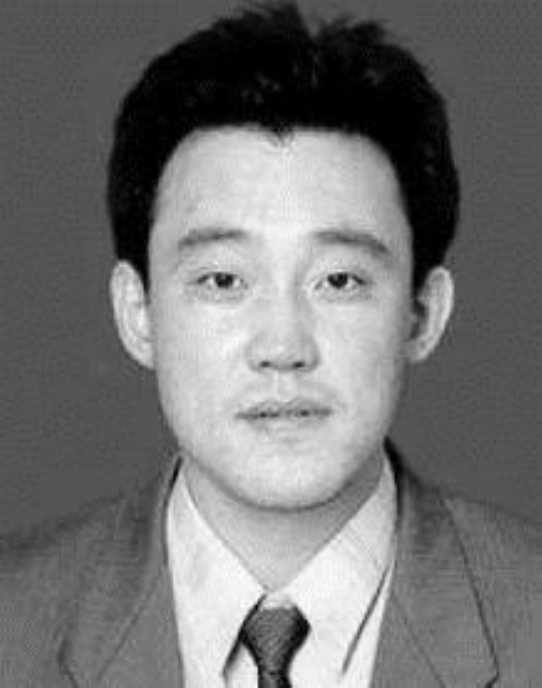}}]{Liang Chu} was born in 1967. He received the B.S., M.S., and Ph.D. degrees in vehicle engineering from Jilin University, Changchun, China. He is currently a Professor and the Doctoral Supervisor with the College of Automotive Engineering, Jilin University. His research interests include the driving and braking theory and control technology for hybrid electric vehicles, which conclude powertrain and brake energy recovery control theory and technology on electric vehicles and hybrid vehicles, theory and technology of hydraulic antilock braking and stability control for passenger cars, and the theory and technology of air brake ABS, and the stability control for commercial vehicle. 

Dr. Chu has been a SAE Member. He was a member at the Teaching Committee of Mechatronics Discipline Committee of China Machinery Industry Education Association in 2006. 
\end{IEEEbiography}
\vspace{-10mm}

\begin{IEEEbiography}[{\includegraphics[width=1in,height=1.25in,clip,keepaspectratio]{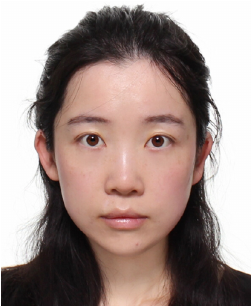}}]{Jingjing Jiang} received the B.Eng. degree in electronic and electrical engineering from the University of Birmingham, Birmingham, U.K., and the B.Eng. degree in measurement, control technology and instrument from the Harbin Institute of Technology, Harbin, China, in 2010, and the M.Sc. degree in control systems and the Ph.D. degree with a thesis on shared control for systems with constraints from Imperial College London, London, U.K., in 2011 and 2016, respectively.

Since September 2018, she has been a Lecturer of Intelligent Mobility/Autonomous Vehicles with the Department of Aeronautical and Automotive Engineering, Loughborough University, Loughborough, U.K. After completing the Ph.D. degree, she was appointed as a Research Associate with the Control and Power Group, Imperial College London. She has authored or coauthored 30 papers in journals and conferences. Her research interests include control design of intelligent systems with constraints, human in the loop, driver assistance control, and autonomous systems.
\end{IEEEbiography}
\vspace{-10mm}

\begin{IEEEbiography}[{\includegraphics[width=1in,height=1.25in,clip,keepaspectratio]{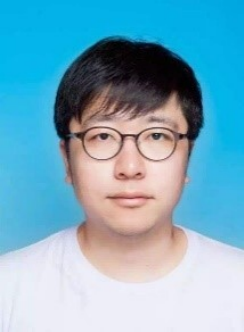}}]{Yuanjian Zhang} (Member, IEEE) received the M.S. in Automotive Engineering from the Coventry University, UK, in 2013, and the Ph.D. in Automotive Engineering from Jilin University, China, in 2018. In 2018, he joined the University of Surrey, Guildford, UK, as a Research Fellow in advanced vehicle control. From 2019 to 2021, he worked in Sir William Wright Technology Centre, Queen’s University Belfast, UK. 

He is currently a Lecturer with the Department of Aeronautical and Automotive Engineering, Loughborough University, Loughborough, U.K. He has authored several books and more than 50 peer-reviewed journal papers and conference proceedings. His current research interests include advanced control on electric vehicle powertrains, vehicle-environment-driver cooperative control, vehicle dynamic control, and intelligent control for driving assist system.
\end{IEEEbiography}
\enlargethispage{-7.7cm}

\end{document}